%% file: main.tex
\documentclass[letterpaper,10pt]{IEEEtran}
\ifCLASSINFOpdf
\else
\fi
\usepackage{amsthm}
\usepackage{amsmath}
\usepackage{graphicx}
\usepackage{multicol}
\usepackage{comment}
\usepackage{multirow} 
\usepackage{epstopdf}
\usepackage{color}
\usepackage{diagbox}
\usepackage{latexsym} 
\usepackage{times}
\usepackage{graphicx}
\usepackage{subfigure}
\usepackage{booktabs}
\usepackage{float}
\usepackage{mathtools, amsfonts, amssymb, amsthm,color,bm} 
\usepackage{dsfont}
\usepackage{cite}
 \usepackage[english]{babel}
\usepackage[figurename=Fig.,font={small,it}]{caption} 
\usepackage{siunitx}
\usepackage{algorithm}
\usepackage{lettrine}
\usepackage{algorithmic} 

\theoremstyle{definition}
\newtheorem{theorem}{\textbf{Theorem}}

\newtheorem{proposition}[theorem]{\textbf{Proposition}}

\usepackage[utf8]{inputenc}
\usepackage[T1]{fontenc}
\usepackage{authblk}
\date{%
}   
 
 \usepackage{xcolor}

\begin{document}
%\title{Real-time Energy Disaggregation at Substation Level via   Dictionary Learning with Partial Labels}
%\title{Uncertainty Bayesian Modeling for Energy Disaggregation with Behind-the-Meter Solar Generation}

\title{
A Decision-Focused Predict-then-Bid Framework for Strategic Energy Storage}

\author[ ]{
\IEEEauthorblockN{Ming Yi, \textit{Member, IEEE,} Yiqian Wu, \textit{Student Member, IEEE,}  Saud Alghumayjan, \textit{Student Member, IEEE}, James~Anderson,  \textit{Senior Member, IEEE,} Bolun Xu, \textit{Member, IEEE}  \thanks{% Paper no.TPWRS-00384-2020. 
	This work was supported in part by the  Data Science Institute in Columbia University and in part by the National Science Foundation under award ECCS-2239046.  (Corresponding author: Ming Yi.)}	\thanks{ Ming~Yi is with Data Science Institute, Columbia University, New York, NY 10027, USA; Saud Alghumayjan, and Bolun Xu  are with Earth and Environmental Engineering, Columbia University, New York, NY 10027, USA; Yiqian Wu, and James Anderson  are with Electrical Engineering, Columbia University, New York, NY 10027, USA. Email:\{my2826, yw3740, saa2244, ja3451, bx2177\}@columbia.edu.}
}}
%E. Farantatos and T. Barik are with Electric Power Research Institute, Palo Alto, CA.
%\{efarantatos, TBarik\}@epri.com,
%} 	 } 
%	 }   
 
%Evangelos Farantatos, \textit{Senior Member, IEEE,} Tapas Barik, \textit{Member, IEEE}

%\affil[1]{ Department of Electrical, Computer, and Systems Engineering, Rensselaer Polytechnic Institute, Troy, NY. }
%\affil[2]{ GEIRI North America, 250 W Tasman Dr, San Jose, CA, 95134.} 

\maketitle 
%With the rapid increase in grid-scale energy storage deployment, it has become crucial for power system operators to accurately predict strategic energy storage behaviors in electricity markets, specifically the timings of charging and discharging. This prediction task is complex due to the intricate interplay of market fluctuations and decision-making processes, constrained by physical power and energy limits. 

\begin{abstract}
This paper introduces a novel decision-focused framework for energy storage arbitrage bidding. % We propose a decision-focused ``predict-then-bid'' strategy that incorporates opportunity value and degradation costs, based on a physical energy storage model.
Inspired by the bidding process for energy storage in electricity markets, we propose a  ``predict-then-bid'' end-to-end method incorporating the storage arbitrage optimization and market clearing models. This is achieved through a tri-layer framework that combines a price prediction layer with a two-stage optimization problem: an energy storage optimization layer and a market-clearing optimization layer.
We leverage the implicit function theorem for gradient computation in the first optimization layer and incorporate a perturbation-based approach into the decision-focused loss function to ensure differentiability in the market-clearing layer. Numerical experiments using electricity market data from New York demonstrate that our bidding design substantially outperforms existing methods, achieving the highest profits and showcasing the effectiveness of the proposed approach.
\end{abstract}

\allowdisplaybreaks
\input{introduction.tex}

\input{Formulation.tex}

\input{Method.tex}

\input{Experiment.tex}

\vspace{-2mm}  
\section{Conclusions}\label{con}
This paper presents a decision-focused framework for energy storage bidding that integrates a physical storage layer and a market-clearing layer into machine learning pipelines. We propose a predict-then-bid strategy and effectively link the dual variable of the SoC transition to bid design. The tri-layer framework, consisting of price prediction, energy storage optimization, and market clearing, enables optimal bidding strategies through end-to-end training. Numerical experiments with real-world NYISO data demonstrate that our framework achieves higher profits than existing methods, validating its effectiveness in enhancing energy storage bidding performance in electricity markets. In future work, we plan to extend our framework to multi-agent energy storage bidding by incorporating multi-locational marginal prices. Additionally, we will explore using the framework to infer sensitivity parameters in price-maker models.

 %\section*{Acknowledgment}
 \vspace{-2mm}  
\bibliographystyle{IEEEtran} 
%\bibliography{./bibfiles/IEEEabrv,./bibfiles/ref,./bibfiles/MengWangPub}
%\bibliography{./bibfiles/IEEEabrv,./bibfiles/ref,./bibfiles/MengWangPub}
\bibliography{IEEEabrv,ref}
 
%\bibliographystyle{IEEEtran} 

%\bibliography{IEEEabrv,ref}

 %\iffalse
% \vspace{0.8in}
%\vspace{-1.5in}
%   \begin{IEEEbiography} [{\includegraphics[width=1.05in,height=1.25in,clip,keepaspectratio]{./Figures/Author_MingYi.pdf}}]	{Ming Yi} (M'23) received the B.E. degree in automation from Harbin Engineering University, Harbin, China, in 2016, and the M.S. degrees in control science and engineering from Harbin Institute of Technology, Harbin, China, in 2018, respectively. 

% He is currently a Ph.D. student in Rensselaer Polytechnic Institute, Troy, NY, USA. His research interests include signal processing, machine learning, power systems monitoring, and high-dimensional data analysis.
% \end{IEEEbiography}

%\fi
% \clearpage
% \newpage
%\vspace{-7in}
%\vspace{-1in}
%\fontsize{10pt}{10pt}\selectfont
%\setlength\columnsep{0.2in}  % Adjust column separation if needed
%\setlength\textwidth{7.1in}  
\input{Appendix}

\end{document}

%% file: introduction.tex
\section{Introduction}

% Energy storage is crucial to advancing the integration of renewable generation by storing the excess energy of renewable generations and moving the supply to periods when it is scarce, e.g., solar generation is low during the nighttime. The past decades have witnessed a remarkable surge of deployment of energy storage in energy systems. The U.S. Energy Information Administration (EIA) reported that the total battery storage capacity in the United States has climbed to 16 GW by the end of 2023 \cite{EIA23}, and it is expected that the total capacity will expand to more than 30 GW by the end of 2024. Facilitated by Federal Energy Regulatory Commission (FERC) Order 841 \cite{FNRC18}, energy storage can participate in all wholesale energy markets in the United States, and price arbitrage has become the most popular service~\cite{zheng2023energy}. On the other hand, distributed energy storage systems also offer increased demand-side flexibility and are becoming key resources participating in demand response programs~\cite{oldewurtel2013framework}.   

Energy storage systems are becoming a mainstream resource in electricity markets, reaching more than 13~GW in California~\cite{CaliforniaStorage2024} and 6~GW in Texas~\cite{TexasStorage2025}. 
To participate in electricity markets, storage resources submit bid offer curves that reflect their willingness to charge or discharge at different prices. These {bidding} curves should capture the storage system’s willingness to charge or discharge based on the market clearing price, accounting for physical operating constraints and anticipated future price opportunities~\cite{zheng2023energy}. Over the past decade, as storage hardware technologies have matured and become more cost-effective, numerous  studies and industry practices have highlighted the importance of effective bidding strategies both for maintaining competitiveness in electricity markets and for facilitating social welfare ~\cite{qin2023role,sioshansi2021energy}.

% Energy storage systems (ESS) have become essential components of modern power grids, offering flexibility to balance supply and demand and supporting the integration of renewable energy sources. Enabled by Federal Energy Regulatory Commission (FERC) Order 841 \cite{FNRC18}, ESS can now participate in all U.S. wholesale electricity markets, providing a wide range of services. Among these, energy arbitrage—charging during low-price periods and discharging during high-price periods—stands out as a vital application \cite{zheng2023energy}. This practice allows storage owners to maximize profits while improving grid efficiency by reducing peak demand and enhancing resilience.

% Energy storage arbitrage can participate in electricity markets through two approaches: self-scheduling and bidding. In self-scheduling, energy storage systems directly determine their charge and discharge schedules based on internal optimization. In contrast, energy storage arbitrage bidding requires storage operators to submit charge and discharge bids to the electricity market operator, where the market-clearing process determines the final charge/discharge amount. This bidding process adds complexity, as storage operators must strategically plan their bids to account for market dynamics and operational degradation costs.

A common approach to energy storage bidding adopts a predict-then-bid framework: a forecast model predicts future prices and the storage operator designs the offer curve based on these forecasts. This decoupling often leads to inefficiencies, because predictors optimized only for accuracy may misalign with the operational objectives of energy storage, resulting in suboptimal charge and discharge strategies.  Accurately predicting electricity prices is especially challenging due to their high volatility, driven by the need to maintain instantaneous supply-demand balance and address local network constraints~\cite{wang2017impact}. Moreover, most forecasting techniques {aim to }minimize the mean squared error which focuses on the accuracy of average prices, {rather than pointwise-in-time differences}. However, energy storage arbitrage hinges on exploiting price differences, making the identification of price peaks and valleys much more critical. 

% Unlike traditional approaches to energy storage self-scheduling, which often rely on Model Predictive Control (MPC) and treat price prediction and self-scheduling optimization as separate stages, the decision-focused framework offers a more integrated approach and has been gaining significant attention. 
%\JA{If possible, keep the paragraph above that is commented out. It provides needed context for the paragraph below. In the paragraph below, it's not clear what you are learning.}

A decision-focused learning approach integrates decision-making criteria into the end-to-end training process, enhancing decision quality by training prediction models that align with operational goals. This approach has demonstrated improved performance when applied to self-scheduling energy storage arbitrage, as shown in our previous work \cite{YAX24}, as well as in other studies~\cite{SXL22}, where it directly generates storage charge and discharge decisions. However, these methods cannot be directly adapted to generate price-quantity offer curves for bid design. Additionally, the market-clearing process, which settles the market based on submitted offer curves and determines dispatch decisions, introduces a feedback loop between bidding strategies and market-clearing outcomes. 

This paper connects price predictions, bidding strategies, and real-time market clearance decisions to generate storage market bids within a decision-focused framework. 
% This paper proposes a novel pipeline for integrating decision-focused learning for storage bid design, fully integrating both the energy storage physical model and market-clearing dynamics.
% To overcome these limitations, effective energy storage bidding design must incorporate price predictions as opportunity costs and account for the degradation costs associated with the physical energy storage model. Fully leveraging the decision-focused approach requires training the prediction model to integrate both the energy storage physical model and market-clearing dynamics, establishing a feedback loop to enhance overall performance.  
The contributions are as follows: 
\begin{itemize}
    \item We propose a tri-layer end-to-end framework to produce optimal arbitrage decisions through bid design, including 1) price prediction based on input features; 2) bid design based on price predictions; and 3) market clearing based on designed bids.
    
   \item We link the storage opportunity dual variable with bidding design in an end-to-end framework, and prove this produces a differentiable implicit layer. % To the best of our knowledge, this specific formulation for bidding design has not been studied before.

   \item We enable efficient backpropagation for the decision-focused loss function through the implicit function theorem and perturbation methods.
    
   \item Numerical experiments on NYISO datasets demonstrate that our approach outperforms benchmark methods.

\end{itemize}

The  paper is organized as follows: Section II reviews the related work, Section III introduces the problem formulation, and Section IV presents our proposed framework and the gradient computation for backward propagation. Simulation results are presented in Section V.%, and Section VI is the conclusion.

\section{Related Work}
\subsection{Learning-Aided Energy Storage Operation}\label{price_pred}

Storage participation in electricity markets can be grouped into {two categories}: self-scheduling and bidding. The self-scheduling option is often limited to small-scale batteries, typically under 1 MW, and yields lower profit expectations. This approach requires dispatch decisions to be made \emph{ahead} of market clearing, reducing flexibility to adjust based on market results. Utility-scale energy storage systems typically submit charge/discharge bids, allowing participants to express varying supply willingness across different price levels~\cite{ruiz2012, hu2021} and embed risk preferences under market uncertainty~\cite{chen2018, zhang2020}. Although economically efficient and theoretically robust, this approach is far more computationally intensive than self-scheduling. For storage systems, bids must capture both operational and opportunity costs, reflecting the future market value of stored energy—an inherently challenging task due to temporal coupling~\cite{shafiee2017, shafiee2019}.

Opportunity cost calculation often employs stochastic dynamic programming (SDP) to account for probabilistic market conditions, albeit at high computational expense~\cite{ZJX22,BZX23}. Model predictive control (MPC) offers an alternative, leveraging rolling-horizon flexibility to manage uncertainty while integrating deterministic constraints~\cite{ADJ16,CZZ17}.
Building on these methods, energy storage operators can submit either energy- or state-of-charge (SoC)-based bids~\cite{zheng2023energy}. Energy-based bids are simpler and align directly with market clearing but may overlook inter-temporal constraints. In contrast, SoC-based bids explicitly incorporate the storage system’s state evolution, enabling more precise operational decisions at the cost of increased market complexity.

\subsection{Decision-Focused Learning}
MPC-based methods typically train machine learning predictors to minimize prediction error, without considering how the predictions will be used in subsequent optimization problems.
Decision-focused learning has emerged as a promising approach to overcome the limitations of traditional MPC-based methods by embedding optimization constraints and objectives directly into supervised learning models, enabling end-to-end training that improves decision quality. Several frameworks have been proposed within this domain, including smart predict-then-optimize approaches \cite{EG22, ELM20}, learning-by-experience methods \cite{BJK24, BBT20}, and black-box differentiable optimizers \cite{PPM20}.

The effectiveness of decision-focused learning has been demonstrated in self-scheduling energy storage arbitrage \cite{YAX24, SXL22, DAK17}. However, as discussed in the Introduction, utility-scale energy storage systems cannot directly charge or discharge. Instead, they must submit market bids, requiring a more complex framework that integrates bid design and a market-clearing layer to determine charge/discharge decisions. Extending a decision-focused framework from self-scheduling  to economic bidding is therefore non-trivial.

Our approach distinguishes itself from existing methods by outputting the dual variable of the SoC evolution constraint, which reflects the marginal opportunity value of energy and informs more strategic bid design. Additionally, we incorporate a market-clearing layer, which calibrates price predictions after price information is revealed to achieve optimal charge/discharge decisions. This approach is particularly beneficial for market participants, such as storage operators, who do not directly control the charge or discharge of storage assets, but instead participate in the market by submitting bids ahead of time. By calibrating these bids after the price is revealed, our framework enhances bidding quality and improves alignment with actual market conditions. Consequently, we generalize decision-focused learning into a more adaptive framework for strategic energy storage market participation.

%% file: Formulation.tex
\section{Problem Formulation}

% We begin by presenting the optimization model for energy storage, followed by the formulation of the learning-enabled bidding problem.
% This paper introduces a decision-focused framework and provides two application scenarios for energy storage arbitrage and predicting energy storage behaviors.  The objective of energy storage arbitrage is to maximize total profit based on future price expectations.  The objective of predicting the energy storage arbitrage behaviors is to predict the timing of the storage charges or discharges, according to historical observed market prices and the storage's past operation profiles. We first outline the energy storage arbitrage decision-making model and then provide a general problem formulation. Then we explain how to fit these two applications into our proposed framework.

\subsection{Bi-Level Storage Bidding Model}

\begin{figure}[t]
  \centering
 { \includegraphics[width = .85\linewidth]{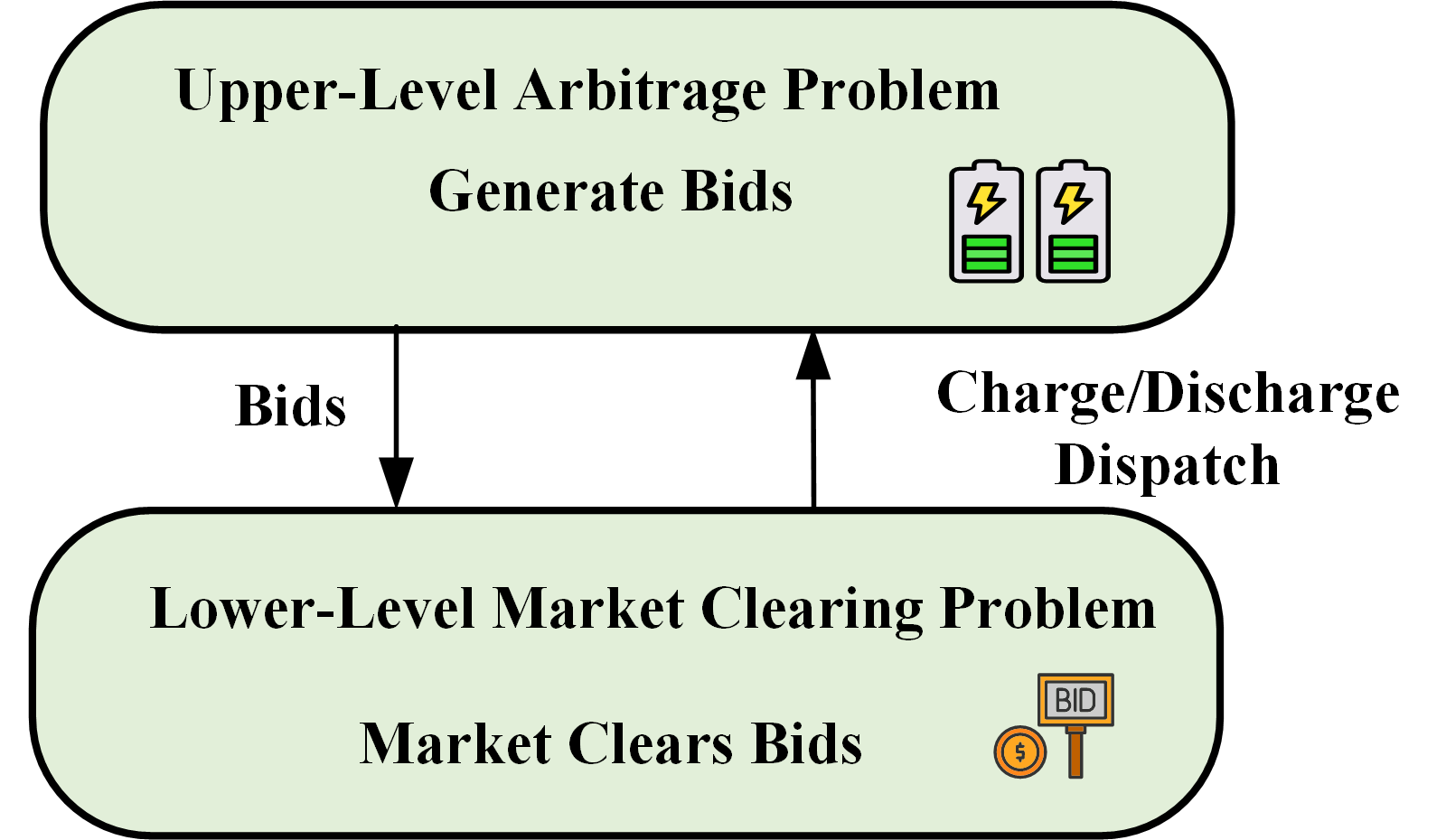}}
 \caption{The bi-level framework illustrates the upper-level energy storage arbitrage problem and the lower-level market clearing process.} \label{Fig: bi-level}
 % \vspace{-1.5em}
\end{figure}

% {\color{red}We consider a bi-level framework to build storage bidding strategy, in which the upper level determines the optimal bids, and the lower level simulates the market clearing process. In the upper level, the storage operator submits bids while accounting for future price uncertainty, the underlying physical model and dispatch decisions from the market clearing problem in the lower level. The bid design can be derived analytically from the storage arbitrage problem, which we will detail in Section IV.A. The lower-level problem clears all submitted bids and determines the corresponding charge and discharge dispatch. Figure~\ref{Fig: bi-level} illustrates the overall bi-level structure. The detailed formulation is given in~\eqref{eq:arb} and will be explained below. The upper-level problem corresponds to \eqref{eq:1a}-\eqref{eq:1e} and the lower-level to \eqref{market_clear}-\eqref{eq:1e}.}

We consider a bi-level framework to build a storage bidding strategy, in which the upper level determines the optimal bids, and the lower level simulates the market-clearing process. Figure~\ref{Fig: bi-level} illustrates the overall bi-level structure. The storage {operator} submits price--quantity segment curves for both discharge and charge, each indexed by \(j \in N\). The discharge/supply curve at time $t$, represented by pairs~\(\{S_{t,j}, P_j\}\), is {monotonically} increasing in $j$ (i.e., \(S_{t,j-1} \leq S_{t,j}\)), indicating the storage will discharge segment \(P_j\) whenever the market price exceeds \(S_{t,j}\). The charge/demand curve \(\{D_{t,j}, B_j\}\) is also monotonic in \(j\), capturing a willingness to charge more as the price drops. To simplify the design process, we assume the segment quantity $P_j$ and $B_j$ are prefixed (for example, equally divide the power rating into $N$ segments). Hence, the decision variables of the upper-level include only $S_{t,j}$ and $D_{t,j}$. {The upper-level (UL) bidding problem then uses future price predictions over a look-ahead period of $T$ to design the bids. The lower-level (LL) problem clears all submitted bids and determines the corresponding charge and discharge dispatch. The bi-level problem is formulated as:}
\begin{subequations}
\label{eq:arb}
{
% \fontsize{9.5pt}{10pt}\selectfont
\begin{align}
   \mathbf{UL:}\;\ \underset{S_t, D_t }{\textbf{max}}  & \ \sum_{t=1}^T {\hat{\lambda}}_{t} (p^*_t - b^*_t) - c(p^*_t, b^*_t) \label{eq:1a}\\[1ex]
          \text{s.t.}
            & \ 0\leq  p^*_t \leq R \,:\quad \underline{u}_t , \bar{u}_t \label{eq:1b}\\
             & \ 0\leq  b^*_t \leq R \,:\quad \underline{v}_t,\bar{v}_t \label{eq:1c}\\
            & \ 0\leq  e_t\leq E \,:\quad \underline{w}_t, \bar{w}_t \label{eq:1d}\\
              &\ e_{t} - e_{t-1}  = -{p^*_{t}}/{\eta} + b^*_{t}\eta \,: \quad \bar{\theta}_t   \label{eq:1e} \\
            \mathbf{LL:}\ p_{t}^*, b_{t}^* &\in   \underset{p_{t}, b_{t}, g_t}{\textbf{argmin}} \  G_t(g_t) + \sum_{j=1}^{N} \left( S_{t,j} p_{t,j} - D_{t,j} b_{t,j} \right) 
            \label{market_clear} \\
    &\quad \ \text{s.t. } % \quad & e_{t,j} - e_{t-1,j} = b_{t,j} \eta - \frac{p_{t,j}}{\eta}    \label{soc_evol}\\
    \ p_t = \sum_{j=1}^N p_{t,j},\; b_t = \sum_{j=1}^N b_{t,j} \label{power_seg}\\
    &\qquad  \ \ \ 0 \leq p_{t,j} \leq P_j\label{ll_supply}  \\
    &\qquad  \ \ \ 0 \leq b_{t,j} \leq B_j\label{ll_demand}  \\
    &\qquad  \ \ \  -e_{t-1} \leq b_{t} \eta - \frac{p_{t}}{\eta} \leq E-e_{t-1} \label{soc_evol} \\
    &\qquad  \ \ \ g_t + \sum_{j=1}^{N} p_{t,j} = L_t +  \sum_{j=1}^{N} b_{t,j} :\quad {\lambda}_t \label{supply-demand}
\end{align}}
\end{subequations}
%\JA{$\lambda(x)$ an input/paramterization of the cost function?  It's also confusing to have $S_t,D_t$ listed as decision variables and not have them appear in the optimization problem.}
where 
\begin{itemize}
    \item $\hat{{{\lambda}}}_t $: Predicted price at time $t$, generated by a pre-trained predictor $g_w(\bm{x})$, where $\hat{{\bm{\lambda}}} = g_w(\bm{x})$. This function  $g(\bm{x})$ represents  future price predictions with input features $\bm{x}$, such as historical load and prices.
    \item $p^*_t, b^*_t$: Optimal discharge and charge decisions at time $t$. Note these optimal decisions should be a function of $S_t$ and $D_t$ based on the lower-level market clearing results.
    \item $c(p^*_t, b^*_t)$: Convex cost function related to storage operation. 
    \item $e_t$: State of charge of storage at time $t$.
    \item $R$: Maximum power rating of the storage system.
    \item $E$: Maximum energy capacity of the storage system.
    \item $\eta$: Efficiency factor of the storage system.
    \item $\bar{\theta}_t$: Dual variable associated with the SoC dynamic constraint representing the marginal opportunity value of the stored energy.

    \item Dual variables associated with \eqref{eq:1b}--\eqref{eq:1d} are indicated following the respective equations, which will later be used for theoretical analysis.
\end{itemize}

\begin{itemize}
    \item $G_t(g_t)$: Aggregated generation cost at time $t$ excluding the target storage.
    \item $g_t$: Net generation at time $t$ excluding the target storage.
    \item $p_{t,j}, b_{t,j}$: Supply and demand quantity for segment $j$ at time $t$.
    \item $P_j, B_j$: Supply and demand capacity for segment $j$.
    \item $L_t$: Total load at time $t$. 
    \item ${\lambda}_t$: Dual variable associated with the supply-demand balance constraint representing market clearing price at time $t$.
\end{itemize}
% Note that we do not include additional time step parameters for simplicity as we consider uniform time steps and normalize all power variables/parameters such as $p^*_t, b^*_t$ with the time step.

In the upper-level  \eqref{eq:1a}-\eqref{eq:1e}, the objective function (\ref{eq:1a}) aims to maximize the expected arbitrage profit including the predicted market revenue and the associated operational cost $c(p^*_t, b^*_t)$. \eqref{eq:1b} and \eqref{eq:1c} enforce the power limits on discharging and charging. \eqref{eq:1d} enforces that the stored energy remains within its capacity limits $E$. \eqref{eq:1e} models the dynamic behavior of the energy storage adjusted by efficiency $\eta$. In the following sections, we denote the constraints \eqref{eq:1b} to \eqref{eq:1e} as the feasibility set    $\mathcal{X}_1$. 
{This bid design can be derived analytically from the storage arbitrage problem~\eqref{eq:1a}-\eqref{eq:1e}, which will be detailed in Section~\ref{bid_design}. }

In the lower-level \eqref{market_clear}-\eqref{supply-demand}, note that each time period~$t\in T$ has a corresponding market clearing problem and, for simplicity, we assume they share a uniform form as we consider uniform time steps and normalize all power variables/parameters such as $p^*_t$, $b^*_t$ with the time step. 
The objective of market clearing is to minimize the aggregated system cost, here, for simplicity, we represent the rest of the market using an aggregated convex objective function including the supply curve from other resources in the system and the congestion impacts (since we target a single storage at a fixed node). 
%A representative market clearing problem over period $t$ is defined as:
%\todo{Use capital letter for parameters}
Therefore, the objective function \eqref{market_clear} minimizes the aggregated system cost $G_t(g_t)$ plus the  net cost of storage charge and discharge, weighted by the corresponding bid prices $S_{t,j}$ and $D_{t,j}$. \eqref{power_seg} models the total charge and discharge power as the sum of all segments. \eqref{ll_supply} and \eqref{ll_demand} limit power supply and demand within predefined segment bounds $P_j$ and $B_j$. \eqref{soc_evol} ensures that charge and discharge operations respect the energy bounds and account for efficiency losses. \eqref{supply-demand} enforces power balance with the associated dual variable ${\lambda}_t$ representing the market clearing price.

\subsection{Price-Taker and Price-Maker Reformulation}
The challenge to solving the bi-level problem \eqref{eq:arb} is the aggregated system cost function $G_t$ is unknown  to the  storage operator, and hence, we cannot solve the problem {exactly}. To solve this issue, we first present two reformulations of LL the arbitrage problem based on price-taker and price-maker assumptions for more convenient integration into our decision-focused learning pipeline.

\textit{Price-Taker Settings.} A price-taker setting assumes that energy storage has a negligible impact on the supply-demand balance and, consequently, on real-time prices.
Historical price data is used to simulate how the energy storage system would have been cleared in the electricity market. In this scenario, we apply a Lagrangian relaxation to \eqref{supply-demand}, incorporating the supply-demand constraint into the objective function. As a result, the lower-level market clearing problem under the price-taker assumption is equivalent to the following storage arbitrage problem:
\begin{align} 
     \underset{p_{t,j},b_{t,j}}{\textbf{max}}  & \quad {\lambda}_{t} \sum_{j=1}^{N}(p_{t,j}-b_{t,j})-\sum_{j=1}^{N}(S_{t,j}p_{t,j}-D_{t,j}b_{t,j})
      \label{eq:clearing1} \\
    \text{s.t.} & \quad \eqref{power_seg}-\eqref{soc_evol} \nonumber
\end{align}

\begin{figure*}[!t]
  \centering
\includegraphics[width=1\linewidth]{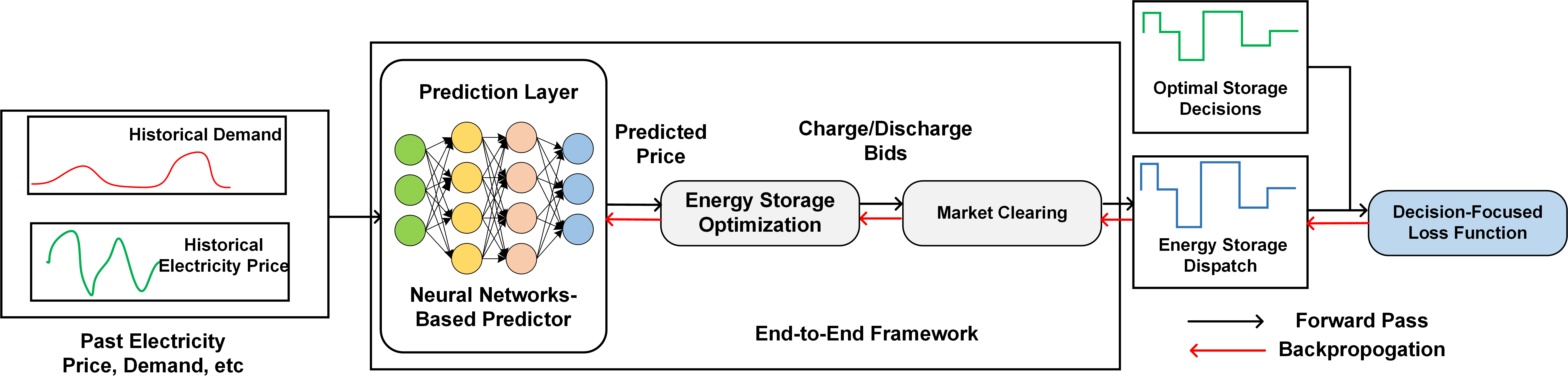}
    	\caption{{The proposed pipeline begins with a neural network-based predictor that forecasts electricity prices from the input features. The optimization layer then leverages the predicted prices, treated as a hidden reward, to solve an optimization problem and compute the dual variable of the state of charge (SoC). This dual variable is linearly transformed into charge/discharge bids. A subsequent market-clearing optimization layer processes these bids to generate charge/discharge dispatches. The algorithm employs backpropagation through both optimization layers to update the predictor's weights.}} \label{Fig: flowchart}
        \vspace{-1em}
\end{figure*}

 % \noindent The constraints in \eqref{soc_seg}-\eqref{soc_seg2} define the feasibility set $\mathcal{X}_2$.
 
\textit{Price-Maker Settings.} A price-maker setting assumes that energy storage will impact the supply-demand balance and, consequently, influence real-time prices. In this case, historical price data cannot be used to simulate how the energy storage system would have been cleared in the electricity market. We still apply a Lagrangian relaxation to \eqref{supply-demand}, embedding the supply-demand constraint into the objective function. Under the price-maker assumption, the market clearing problem becomes equivalent to the following storage arbitrage problem:
 \begin{align}
     \underset{p_{t,j},b_{t,j}}{\textbf{max}}  & \ {\lambda}_{t}(p_t, b_t) \sum_{j=1}^{N}(p_{t,j}-b_{t,j})-\sum_{j=1}^{N}(S_{t,j}p_{t,j}-D_{t,j}b_{t,j})
     \label{eq:clearing2} \\
     \text{s.t.} &\ \eqref{power_seg}-\eqref{soc_evol} \nonumber 
 \end{align}
The key in the price-maker reformulation is to model the market clearing price as a function of charge and discharge power with a sensitivity function as:
\begin{align}\label{sensitivity_model}
    \lambda_{t}(p_t, b_t) = {\lambda}_{t}-f(p_t- b_t)
\end{align}
where ${\lambda}_{t} = \mathrm{d} G_t(L_t)$ represents the market clearing price without our storage, which is the marginal cost of the aggregated cost function to supply the demand $L_t$. $f(\cdot)$ is a monotonic increasing function with $f(0)=0$ representing the impact of the storage action over the market price. Ideally, ${\lambda}_{t}-f(p_t- b_t)$ should be equivalent to $\mathrm{d} G_t(L_t+b_t-p_t)$ representing the change in the system marginal cost, yet the proposed formulation highlights that $G_t$ is not fully observable. 

%\JA{is $d$ a constant or an infinitesimal, in which case use $\mathrm d$}

% \noindent For a linear price-sensitive model, the relationship is given by $\lambda_{t}(p_t, b_t) = {\lambda}_{t}-\alpha (p_t- b_t)$, where $\alpha$ represents the price sensitivity parameter. 

\subsection{Problem Statement and Learning Pipeline}

We now propose the main decision-focused framework using two-stage optimization. In stage one, predicted prices generate charge/discharge bids via an energy storage optimization model. In stage two, these bids feed into a market-clearing optimization. A loss function minimizes the difference between decisions from predicted bids and optimal decisions, backpropagating through both stages to update the price predictor. The end-to-end training policy maps the input features to the energy storage dispatches $(\bm{p}^*, \bm{b}^*)$ as follows:
\begin{equation}\label{eq:mapping_nn}
\bm{x} \mapsto (\bm{p}^*, \bm{b}^*) = f_{\mathrm{L}}\left(f_{\mathrm{U}}\left(g_w\left(\bm{x}\right)\right)\right)
\end{equation}
\noindent where $\bm{x}$ represents input features and $g_w\left(\bm{x}\right)$ is the neural network model for price prediction. $f_{\mathrm{U}}(\cdot)$ represents the UL problem, which maps the price prediction to the bids $(S_{t,j}, D_{t,j})$, and $f_{\mathrm{U}}(\cdot)$ represents the LL problem which maps the bids and the actual price to the storage dispatch $(\bm{p}^*, \bm{b}^*)$, where the LL problem  can use either the price-taker or price-maker formulation described in Section III.B.

Note that the final output from this learning pipeline are the storage bids that the storage operator will eventually submit to the market, while the storage decision is used for defining the loss function.

\textbf{Training Dataset.}
 We collect input features and  build a dataset as follows:
\begin{equation}
\mathcal{D}=\{(\bm{x}^1,\bar{\bm{y}}^1), (\bm{x}^2, \bar{\bm{y}}^2),...,(\bm{x}^n, \bar{\bm{y}}^n)\}
\end{equation}
where $\bm{x}$ includes historical electricity prices, demand profiles, and other contextual information.  $\bar{\bm{y}}$ denotes the optimal energy storage decisions ($\bar{\bm{p}}$, $\bar{\bm{b}}$). $n$ is the number of data samples. $\{\bar{\bm{y}}^1,\bar{\bm{y}}^2,...,\bar{\bm{y}}^n \}$ represents the set of optimal charge/discharge decisions obtained from historical real-time prices, which can be generated by solving:
\begin{align}\label{optimal_solution}
   \underset{p_t, b_t \in \mathcal{X}_1}{\textbf{max}}  & \quad\sum_{t=1}^T {{\lambda}}_{t} (p_t - b_t) - c(p_t, b_t).
\end{align}

%% file: Method.tex
\section{Methodology}

% \begin{figure*}[!ht]
%   \centering
% \includegraphics[width=1\linewidth]{Figures/FlowChart-Bidding.png}
%     	\caption{{The proposed pipeline begins with a neural network-based predictor that forecasts electricity prices from the input features. The optimization layer then leverages the predicted prices, treated as a hidden reward, to solve an optimization problem and compute the dual variable of the state of charge (SoC). This dual variable is linearly transformed into charge/discharge bids. A subsequent market-clearing optimization layer processes these bids to generate charge/discharge dispatches. The algorithm employs backpropagation through both optimization layers to update the predictor's weights.}} \label{Fig: flowchart}
%         \vspace{-1em}
% \end{figure*}

% Model Predictive Control is a standard strategy for energy storage arbitrage. However, designing energy storage bidding requires formulating charge/discharge bids based on degradation and opportunity costs. 
We propose a ``predict-then-bid'' approach, linking the bidding design to the dual variable of the SoC evolution constraint~\eqref{eq:1e}. The framework has three components: a prediction layer that forecasts electricity prices using historical features (e.g., DAP and RTP), an energy storage optimization layer using these forecasts, and a market clearing layer that simulates dispatch based on submitted bids. Two-stage optimization layers support both forward and backward passes, ensuring weight alignment and updating the prediction layer. An overview is provided in Fig. \ref{Fig: flowchart}.
% Our pipeline integrates two-stage optimization layers. These layers support both forward and backward passes, ensuring alignment of weights and enabling updates in the prediction layer. An overview of the proposed approach is illustrated in Fig. \ref{Fig: flowchart}. In the following, we first introduce the MPC-based energy storage bidding design, followed by details of the prediction and optimization layers. Finally, we describe the backpropagation process within the framework.

%\subsection{Two-Stage Optimization Layers} \label{bid_design}
\subsection{Bidding Design Based on Opportunity Value} \label{bid_design}

For the online operation of energy storage, a rolling horizon approach is used to predict future electricity prices over a finite time horizon from $t$ to  $T$. To design the charge/discharge bid at time 
$t$, the energy storage system must account for the marginal cost of opportunity value and degradation cost. Specifically, the opportunity value represents the objective function with respect to the optimal charge/discharge decisions from 
$t+1$ to $T$. To model the marginal opportunity value with respect to charge and discharge, it is first necessary to express the opportunity value as a function of the state of charge at time $t$. The marginal cost with respect to charge and discharge then follows from the chain rule. This leads to Proposition 1.

\begin{proposition}\label{prop:two} 
\emph{Marginal cost of the opportunity value function.}  Consider the predict-then-bid energy storage arbitrage problem given in \eqref{eq:1a}–\eqref{eq:1e}, assuming that both the primal and dual problems have non-empty feasible sets.  The marginal cost of the opportunity value function $V_{t+1}$  with respect to the state of charge at time 
$t$ is defined by the dual variable of the SoC transition equation \eqref{eq:1e} at time 
$t+1$: 
\[
\frac{\partial V_{t+1}}{\partial e_t}  = \bar{\theta}_{t+1}.
\]
The marginal costs with respect to discharge  $p_t$ and charge $b_t$ at time $t$ are  
\begin{itemize}
    \item  $\frac{\partial V_{t+1}}{\partial p_t} =   \frac{\partial V_{t+1}}{\partial e_t} \cdot \frac{\partial e_{t}}{\partial p_t} = -\bar{\theta}_{t+1} \cdot \frac{1}{\eta},$
    \item  $\frac{\partial V_{t+1}}{\partial b_t} =   \frac{\partial V_{t+1}}{\partial e_t} \cdot \frac{\partial e_{t}}{\partial b_t} = \bar{\theta}_{t+1} \cdot \eta.$
\end{itemize}  
\vspace{2mm}
\noindent where $\bar{\theta}_{t+1}$ denotes the dual variable corresponding to the SoC dynamic constraint in equation \eqref{eq:1e} at time $t+1$.
\end{proposition}

\noindent The proof is provided in the Appendix. The value of $\bar{\theta}_{t+1}$ is dependent on SoC $e_t$. For bidding design, it is crucial to account for opportunity costs across different SoC levels. Thus, we discretize the state of charge $e_{t,j}$ from $0$ to $E$ in $N$ segments and compute the corresponding segment of $\theta_{t+1, j}$.

The discharge bid incorporates both discharge costs and opportunity costs:
\begin{subequations}\label{eq:bid}
\begin{align}
    \hat{S}_{t,j}=  \frac{\partial c(p_t,b_t)}{\partial p_t}+\bar{\theta}_{t+1, j}  \cdot\frac{1}{\eta} 
    \label{eq:dis_bid}
\end{align}

\noindent where ${\partial c(p_t,b_t)}/{\partial p_t} = C$ when energy storage has linear cost term $c(p_t,b_t) = C p_t$.  Similarly, the charge bid considers both charge costs and opportunity costs:
 \begin{equation}
     \hat{D}_{t,j}= \frac{\partial c(p_t,b_t)}{\partial b_t}+ \bar{\theta}_{t+1, j} \cdot \eta
     \label{eq:ch_bid}
 \end{equation}
\end{subequations}

%\todo{I removed the remark here as I don't know what this means/serves}
%\noindent  \textit{Remark.}  After computing an $N$-segment bid, the storage operator can further reduce the number of segments by averaging them to comply with specific market rules.

%Discharging  results in a loss of opportunity value, reflected as a negative value in \eqref{dual1}. We incorporate the marginal opportunity value into the discharge bid design in \eqref{eq:dis_bid}, ensuring that the combined cost accurately represents the minimum price at which energy storage is profitable to discharge, preventing underbidding. Conversely, charging increases opportunity value. When the price is below the sum of the marginal opportunity value and the degradation cost, charging becomes profitable. Therefore, the charge bid is set as the sum of the marginal opportunity value and the marginal degradation cost in \eqref{eq:ch_bid}.

\subsection{Differentiation  of Energy Storage Optimization Layer}
After showing that the optimal storage bids depend on the dual $\bar{\theta}_{t+1, j}$, our next step is to develop a differentiable layer to integrate the dual variable into the learning pipeline. We apply the implicit function theorem, as first introduced in OptNet\cite{amos2017}, to compute the gradient of the dual variable of SoC dynamic constraint with respect to price prediction. While many existing works focus on computing the gradient of primal solutions~\cite{SX24,DAK17}, few have explored the gradients of dual variable with respect to price predictions.
Our objective is to compute the gradient of the dual variable for the SoC transition with respect to the price prediction, ${\partial \bm{\theta}}/{\partial \hat{\pmb{\lambda}}}$, to enable backpropagation of the decision-focused loss. $ \bm{\theta} = [\theta_1, \theta_2... \theta_{N}] \in \mathbb{R}^{N}$ denotes the SoC dual variables associated with the charge/discharge bid formation. Proposition 2 formalizes the gradient computation of the dual variable using the implicit function theorem.

\begin{proposition}  
\emph{Differentiation of the dual variable of SoC transition.}  Consider the Lagrangian associated with the optimization problem \eqref{eq:1a}–\eqref{eq:1e}. Let \(\bm{\mu}\) denote the dual variables corresponding to all inequality constraints. $\bar{\bm{\theta}}_t$ denotes the dual variable of equality constraints of SoC transition. The Karush-Kuhn-Tucker (KKT) conditions define the function:
\[
\mathcal{F}(\bm{p}, \bm{b}, \bm{e}, \hat{\bm{\lambda}}, \bar{\bm{\theta}}_t, \bm{\mu}) = 0.
\]

\noindent Based on the implicit function theorem and chain rule, differentiating the KKT system with respect to \(\hat{\bm{\lambda}}\) gives:
\[
\frac{\partial \mathcal{F}}{\partial \hat{\bm{\lambda}}} 
+ \frac{\partial \mathcal{F}}{\partial \bar{\bm{\theta}}_t} 
\frac{\partial \bar{\bm{\theta}}_t }{\partial \hat{\bm{\lambda}}} = 0.
\]

\noindent Rearrange this equation to obtain the derivative:
\[
\frac{\partial \bar{\bm{\theta}}_t }{\partial \hat{\bm{\lambda}}} = 
- \left( \frac{\partial \mathcal{F}}{\partial \bar{\bm{\theta}}_t} \right)^{-1} 
\frac{\partial \mathcal{F}}{\partial \hat{\bm{\lambda}}}.
\]
\end{proposition}  
%\begin{proof}
\noindent The detailed proof is provided in the Appendix. After computing the matrix \({\partial \bar{\pmb{\theta}}_t}/{\partial \hat{\pmb{\lambda}}} \in \mathbb{R}^{T \times T}\), we extract its first row and repeat this process for each of the \(N\) initial SoC levels. The resulting gradients from all segments are then stacked to construct the final Jacobian matrix \({\partial \pmb{\theta}}/{\partial \hat{\pmb{\lambda}}} \in \mathbb{R}^{N \times T}\).

\subsection{Perturbed Decision-Focused Loss Function}

Given the predicted bids, we aim to ensure that the resulting charge and discharge dispatch is as close as possible to the optimal dispatch. To achieve this, we employ the Fenchel-Young loss \cite{BMN20}, and define the perturbed loss function as follows:
\begin{subequations}
\begin{equation} \label{loss_expect}
\resizebox{0.49\textwidth}{!}{$
\begin{aligned}
& \mathcal{L}^{DF}_\epsilon \\ & =   \mathbb{E}\Big[\underset{p_{t,j},b_{t,j} \in \mathcal{X}_2}{\textbf{max}} \, {\lambda}_{t} \sum_{j=1}^{N}(p_{t,j}-b_{t,j})-\sum_{j=1}^{N}(\Tilde{S}_{t,j}p_{t,j}- \Tilde{D}_{t,j}b_{t,j}) \Big]\\&
\quad - \left( {\lambda}_{t} \sum_{j=1}^{N}(\bar{p}_{t,j}-\bar{b}_{t,j})-\sum_{j=1}^{N}(\hat{S}_{t,j}\bar{p}_{t,j}-\hat{D}_{t,j}\bar{b}_{t,j}) \right)
\end{aligned}
$}
\end{equation}
\noindent where recall that $p_{t,j},b_{t,j}$ are the decision variable, ${\lambda}_{t}$ is the historical price  (or price sensitivity functions ${\lambda}_{t}(p_{t,j},b_{t,j})$ in the price-maker case),  $\bar{p}_{t,j},\bar{b}_{t,j}$ are the hindsight optimal solution generated using historical price data as in \eqref{optimal_solution}, $(\hat{S}_{t,j}, \hat{D}_{t,j})$ are the predicted bids based on the aforementioned dual approach as in \eqref{eq:bid}, and   $(\Tilde{S}_{t,j}, \Tilde{D}_{t,j})$ are the perturbed predicted bids 
\begin{align}
    \Tilde{S}_{t,j}&=  \frac{\partial c(p_t,b_t)}{\partial p_t}+(\bar{\theta}_{t+1, j}+\epsilon Z )  \cdot\frac{1}{\eta} 
    \label{eq:dis_bidp}\\
    \Tilde{D}_{t,j}&= \frac{\partial c(p_t,b_t)}{\partial b_t}+ (\bar{\theta}_{t+1,j}+\epsilon Z )\cdot \eta
     \label{eq:ch_bidp}
 \end{align}
 \end{subequations}
\noindent where ${Z}$ is additive Gaussian noise $\mathcal{N}({0}, 1)$.  $\epsilon >0$ is a scaling parameter. $\mathbb{E}[\cdot]$ denotes the expectation over the random variable ${Z}$. 

%We use $F_{2,\epsilon}(\bar{\bm{\theta}}_{t+1})$ to denote perturbed function in \eqref{eq:pertub}.

The reason we need to add perturbation to the bids prediction is to improve the differentiability, as the bids are piece-wise linear.  Small changes to the charge or discharge bids typically do not affect the final dispatch decisions; In rare cases, such variations may lead to abrupt changes in the decisions. Consequently, the gradient is almost zero everywhere, or it becomes undefined at points where decisions jump.  This discontinuity introduces computational difficulties when applying backpropagation to update the weights of neural network models.  We employ the perturbation approach \cite{BBT20} to address this challenge by introducing additive noise to the dual variable $\theta_{t+1, j}$. 

Correspondingly, the resulting perturbed market clearing model is defined as:
\begin{equation}
\begin{aligned}
% \Tilde{S}_t, \Tilde{D}_t \mapsto  \mathbb{E}\Big[ 
\underset{{p_t, b_t \in \mathcal{X}_2}}{\textbf{max}}\,  &  {\lambda}_{t} \sum_{j=1}^{N}(p_{t,j}-b_{t,j})-\sum_{j=1}^{N}(\Tilde{S}_{t,j} 
p_{t,j}-\Tilde{D}_{t,j}b_{t,j})   
% \Big] 
\label{eq:pertub}\\
\end{aligned}
\end{equation}
\noindent For convenience or later deviations, we use $F_{2,\epsilon}(\bar{\bm{\theta}}_{t+1})$ to denote \eqref{eq:pertub}. The perturbed optimal solution is denoted as  $(\bm{p}_\epsilon^*, \bm{b}_\epsilon^*)$. The differentiability of the perturbed loss function follows from the differentiability of $F_{2,\epsilon}(\bar{\bm{\theta}}_{t+1})$.  The result is formalized in Proposition~3.

\begin{proposition} \label{prop:three}
\emph{Differentiability of loss function with perturbed bid.}  Assume the noise $\bm{Z}$ follows a Gaussian distribution with density $\vartheta(\bm{Z}) \propto \exp(-\psi(\bm{Z}))$.  For notational simplicity, define $\bm{y}_{\epsilon}^* = \bm{p}_\epsilon^*-\bm{b}_\epsilon^*$.
 \begin{itemize}
 \item The perturbed loss function $\mathcal{L}^{DF}_\epsilon$ is twice differentiable, and 
$\nabla_{\bar{\bm{\theta}}_{t+1}} F_{2,\epsilon}(\bar{\bm{\theta}}_{t+1})=\bm{y}_{\epsilon}^*(\bar{\bm{\theta}}_{t+1}) =\mathbb{E}[ y^*(\bar{\bm{\theta}}_{t+1}+\epsilon \bm{Z})]=\mathbb{E}[F_2(\bar{\bm{\theta}}_{t+1}+\epsilon \bm{Z})\nabla_Z \psi(\bm{Z})/\epsilon]$ 
\item $\bm{y}_{\epsilon}^*$ is differentiable, and  $J_{\theta}\bm{y}_{\epsilon}^*=\mathbb{E}[F_2(\bar{\bm{\theta}}_{t+1}+\epsilon \bm{Z})(\vartheta(\bm{Z})\vartheta^{\top}(\bm{Z})-\nabla^2\vartheta(\bm{Z}) ) /\epsilon]$
\end{itemize}  

\end{proposition}

\noindent The proof is provided in the Appendix. Proposition~\ref{prop:three} establishes that the perturbed loss function is differentiable and further demonstrates that the perturbed optimal solution is differentiable with respect to the bids.

%\noindent where $\lambda_t(p_t, b_t)$ is a function of charge/discharge decisions $p_t, b_t$.

\begin{algorithm} [!t]
\caption{Decision-Focused Energy Storage Bidding} 
\begin{algorithmic}[1]  \label{alg1}

\REQUIRE The rolling horizon window $T$; energy storage parameters, including the cost function $c(p_t, b_t)$, power rating $P$, efficiency $\eta$, and energy capacity $E$; the pre-trained price forecasting model; the maximum number of training epochs $M_{\max}$; the training dataset $\mathcal{D} = \{(\bm{x}^1, \bar{\bm{y}}^1), (\bm{x}^2, \bar{\bm{y}}^2), \dots, (\bm{x}^n, \bar{\bm{y}}^n)\}$; and the testing dataset $\hat{\mathcal{D}} = \{\hat{\bm{x}}^1, \hat{\bm{x}}^2, \dots, \hat{\bm{x}}^n\}$.

\textbf{Initialization:} Initialize the weights $\bm{w}$ of the predictor $g_w(\bm{x})$ using the pre-trained model.
 
\textbf{Training stage:}
 \FOR{Epochs < $M_{\max}$ }
 \FOR{Each batch in $\mathcal{D}$ }
 \STATE{The predictor outputs  the electricity price: $\hat{\bm{\lambda}}=g_w(\bm{x})$; }
  \STATE{Forward pass $\hat{\bm{\lambda}}$  to the optimization problem \eqref{eq:1a}-\eqref{eq:1e} to compute the dual variable $\bar{\bm{\theta}}_t$ of the SoC transition;}
  \STATE{Compute the perturbed discharge and charge bids  using \eqref{eq:dis_bidp} and \eqref{eq:ch_bidp};}
 \STATE{Compute the perturbed loss $\mathcal{L}^{DF}_\epsilon$ by equation (\ref{eq:ptb_loss})};
  \STATE{Compute the gradient using  (\ref{eq:total}) and perform backpropagation to update the predictor's weights $\bm{w}$};
 \\

\ENDFOR
\ENDFOR

%\STATE{\bm{Y}=\mathcal{H}^{\dagger}\bm{X}}

\textbf{Testing stage:}

\STATE{The predictor outputs the electricity price: $\hat{\bm{\lambda}}=g_w(\hat{\bm{x}})$; }

 \STATE{Forward pass $\hat{\bm{\lambda}}$  to the optimization problem \eqref{eq:1a}-\eqref{eq:1e} to compute the dual variable $\bar{\bm{\theta}}_t$ of the SoC transition;}

 \STATE{Compute the discharge and charge bids $\hat{S}_t, \hat{D}_t$ using \eqref{eq:dis_bidp} and \eqref{eq:ch_bidp};}
%\STATE{Compute the  by ($\ref{hruncertaintyindex}$)}

%\STATE{Compute the uncertainty index $U_{\textrm{index}}$ by  (\ref{Uncertaintyindex2})}
%\STATE{Compute the predictive mean and uncertainty index by  and , respectively}
\RETURN
Predicted energy storage bids  $(\hat{S}_t, \hat{D}_t)$. %The uncertainty index $U_{\textrm{index}}$.

\end{algorithmic} 
\end{algorithm}

\subsection{Algorithm Implementation}
We now assemble the final algorithm for training and testing the proposed framework.
Since the exact computation of the expectation in \eqref{loss_expect} is intractable, we approximate it using Monte Carlo sampling. The perturbed loss function can thus be estimated as:
\begin{equation}\label{eq:ptb_loss}
% \resizebox{0.49\textwidth}{!}{$
\begin{aligned}
 \mathcal{L}^{DF}_\epsilon  \approx   &\frac{1}{K} \sum_{m=1}^K\bigg(\underset{p_{t,j},b_{t,j} \in \mathcal{X}_2}{\textbf{max}} {\lambda}_{t} \sum_{j=1}^{N}(p_{t,j}-b_{t,j})\\ &-\sum_{j=1}^{N}(\Tilde{S}_{t,j}^{(m)}p_{t,j}- \Tilde{D}^{(m)}_{t,j}b_{t,j})  \bigg) \\&
- \left({\lambda}_{t} \sum_{j=1}^{N}(\bar{p}_{t,j}-\bar{b}_{t,j})-\sum_{j=1}^{N}(\hat{S}_{t,j}\bar{p}_{t,j}-\hat{D}_{t,j}\bar{b}_{t,j}) \right)
\end{aligned}
% $}
\end{equation}
%Also note that we do not need to perturb the $\hat{\lambda}_t$ for the second term because that term is continious. 
\noindent where  $K$ represents the total number of Monte Carlo samples used for the approximation.

%As the optimization layers are located at the end of a pipeline,  our perturbed loss function can simplify
%differentiation as we can compute  $\frac{\partial \mathcal{L}^{DF}_\epsilon}{\partial \bm{\theta}}$ directly without have additional computing $\frac{\partial \mathcal{L}^{DF}_\epsilon}{\partial \bm{y}^*}\frac{\partial \bm{y}^*}{\partial \bm{\theta}}$. 

%\cite{BMN20} shows that the perturbed loss function is differentiable.\\

%\noindent 
%and it is so-called Fenchel-Young Loss \cite{BMN20}. We can define the Fenchal conjugate $\Omega^*(\theta)= \mathbb{E}[\textbf{max}_{p_t, b_t}  \{\sum_{t=1}^T (\hat{\lambda}_{t} +\epsilon Z) y_t-C p_t\}]- ({\sum_{t=1}^T \hat{\lambda}_{t} \bar{y}_t-C \bar{p}_t)$. The original function is $\Omega(\bar{p},\bar{b})$. Then $ \mathcal{L}_\epsilon^{FY}(\hat{\lambda},\bar{p},\bar{b} )+\Omega(\bar{p},\bar{b})$ is positive by Fenchel's inequality, convex and the minimum 0 is reached at $\hat{\lambda}$ if and only if $f_\epsilon(\hat{\lambda})=\bar{p},\bar{b}$.
 %The Fenchel-Young loss has some very nice properties, we can easily get the 

Backpropagation computes the gradient of the loss function with respect to the neural network weights $\bm{w}$ by applying the chain rule:
\begin{align}
    \frac{\partial \mathcal{L}^{DF}_{\epsilon}}{\partial \bm{w}}&=    \frac{\partial \mathcal{L}^{DF}_\epsilon}{\partial \bm{\theta}} \frac{\partial \bm{\theta}}{\partial \hat{\bm{\lambda}}}  \frac{\partial \hat{\bm{\lambda}}}{\partial \bm{w}} \label{eq:total}
\end{align} 
where the gradient of the perturbed loss function with respect to $\bm{\theta}$ is given as:
\begin{equation}
 \frac{ \partial \mathcal{L}_\epsilon^{DF}}{\partial \bm{\theta}}= \left( \frac{\bar{\bm{p}}}{\eta} - \bar{\bm{b}} \eta \right) - \left( \frac{\pmb{p}_\epsilon^*}{\eta}  - \pmb{b}_\epsilon^* \eta  \right)  .
\end{equation}
\noindent The second term ${\partial \bm{\theta}}/{\partial \hat{\bm{\lambda}}}$ can be calculated according to results from Proposition~2. The third term ${\partial \hat{\bm{\lambda}}}/{\partial \bm{w}}$ represents the Jacobian matrix of the price prediction model with respect to the weights  $\bm{w}$, which can be efficiently computed using automatic differentiation provided by deep learning frameworks such as PyTorch or TensorFlow \cite{PGS17}. 

\textbf{Algorithm 1} shows the full decision-focused energy storage bidding pipeline. Our method can be trained from random initialization; but we recommend a pre-trained electricity price forecasting model that minimizes forecasting error for optimal performance.  % In our implementation, the neural network is first trained to minimize the forecasting error, and the resulting model is then used to initialize the decision-focused training process.

\noindent %\textit{Remark.}  Bidding at an hourly timescale for energy storage  leads to lower profits than bidding at a higher temporal resolution. This outcome arises because the variance of the hourly average price is reduced by the averaging process. Proposition 4 provides a rigorous modeling of the variance reduction in hourly prices.
%\begin{proposition} \emph{Variance reduction in hourly average electricity prices.}
%\label{prop:hourly_avg_var}
%Let 
%\[
%   \zeta_{h,1},\, \zeta_{h,2},\, \dots,\, \zeta_{h,n}
%\] 
%be random variables representing high-resolution electricity prices within hour \(h\).  $n$ denotes the number of segments.
%Assume \(\zeta_{h,i}\) has the finite mean \(\mu_\zeta\) and the finite variance \(\sigma^2\).
%\begin{enumerate}
%    \item Each \(\zeta_{h,i}\) has the same finite mean \(\mu_\zeta\) and the same finite variance \(\sigma^2\);
%    \item They are not all perfectly (positively) correlated; i.e., not every pair \(\zeta_{h,i}, \zeta_{h,j}\) has correlation \(=1\).
%\end{enumerate}

%\noindent Define the hourly average
% \[
%   \overline{\zeta}_h 
%   \;=\; 
%   \frac{1}{n} \sum_{i=1}^n \zeta_{h,i}.
% \]
% Then the variance of hourly average price reduces
% \[
%   \mathrm{Var}\bigl(\overline{\zeta}_h\bigr)
%   \;\le\;
%   \sigma^2,
% \]
% with equality if and only if all \(\zeta_{h,i}\) are perfectly correlated.  
% In typical cases, we have
% \[
%   \mathrm{Var}\bigl(\overline{\zeta}_h\bigr) 
%   \;<\;
%   \mathrm{Var}(\zeta_{h,i}) 
%   \;=\;
%   \sigma^2.
% \]
% \end{proposition}

% The proof is included in the Appendix, where readers can find a detailed explanation.

%% file: Experiment.tex
\vspace{-2mm}  
\section{Experiments}
The experiments are conducted using PyTorch~\cite{PGM19} and the PyEPO package~\cite{TK22} on a desktop equipped with a 3.0 GHz Intel Core i9 processor, an NVIDIA 4080 GPU with 16 GB of memory, and 32 GB of RAM. The differentiable optimization layer is implemented using the cvxpylayer package~\cite{AAB19}. The optimization problems are solved by the CVXPY package~\cite{SS16}.
The energy storage arbitrage model is configured with the following parameters: power rating $P = 0.5 \, \text{MW}$, efficiency $\eta = 0.9$, storage capacity $E = 1 \, \text{MWh}$ and initial SoC $e_0 = 0.5 \, \text{MWh}$.  The prediction horizon is set to 24 hours ($T = 24$) and the number of Monte Carlo samples is $K = 1$.

% In this paper, we employ a  convolutional long short-term memory (ConvLSTM) neural network. The model architecture consists of three 1D convolutional layers with 64, 128, and 64 filters, respectively, and a kernel size of 3. Each convolutional layer is followed by a ReLU activation function, with max pooling applied after each layer. The CNN output is reshaped and passed into two bidirectional LSTM layers with 100 hidden size. A dropout layer with a rate of 0.5 is applied after each LSTM layer to prevent overfitting. The final output is generated through a fully connected layer that maps to the target output size.  The model is trained using the Adam optimizer with a learning rate $10^{-4}$. The input features consist of a 24-hour window for historical real-time price, day-ahead price, and load data, and the neural network model predicts the subsequent 24-hour real-time price.  These input features are commonly used for training price forecasting models. To enhance the training samples, a rolling horizon window with a step size of one hour is applied. The data is sampled at an hourly resolution, and the dataset comprises both input features and corresponding target decision values.

We employ a convolutional LSTM (ConvLSTM) network. The architecture features three 1D convolutional layers with 64, 128, and 64 filters, respectively, each using a kernel size of 3. Each convolutional layer is followed by a ReLU activation and a max pooling layer. The feature representation is reshaped and passed to a two-layer bidirectional LSTM, with each layer having a hidden size of 100 and a dropout rate of 0.5. A final fully connected layer maps the LSTM output to the target prediction. The model is trained using the Adam optimizer with a learning rate of $10^{-4}$. The dataset is sampled at hourly resolution, and each input consists of a 24-hour window of historical real-time prices, day-ahead prices, and load data. The network is trained to predict real-time prices for the subsequent 24 hours. A rolling horizon strategy with a one-hour step size is used to augment the training set.

\vspace{-3mm}  
\subsection{Energy Storage Bidding Under Price-Taker Setting }\label{exp:taker}

% We conduct experiments on energy storage bidding, where the energy storage system designs bids based on opportunity costs and degradation costs. The charge/discharge dispatch is determined after the real-time price is revealed. We assume a price-taker setting, and the charge/discharge decisions do not influence the price. 
We start with a price-taker case study. The price forecasting window spans 24 hours and is updated hourly, enabling the bid design to adapt dynamically. The electricity price and load data are sourced from the New York Independent System Operator (NYISO), utilizing data from 2017–2018 for training and 2019 for testing. Historical data for Real-Time Price (RTP), Day-Ahead Price (DAP), and load are collected at an hourly resolution. The dataset contains both input features and corresponding optimal decision values, where the optimal decisions are obtained by solving \eqref{optimal_solution} using the ground truth real-time prices.

In the testing stage, the profit is computed as follows:
\begin{equation}
    \textstyle\sum_{t=1}^T {\lambda}_{t} (p_t^*-b_t^*)-c(p^*_t,b_t^*)
\end{equation}
\noindent where ${\lambda}_t$ represents the actual real-time price at time $t$; $p_t^*$ and $b_t^*$ denote the discharge and charge amounts, respectively, after market clearing.

\begin{figure}[!ht]
  \centering
   \subfigure[]{ \includegraphics[trim = 3mm 0mm 3mm 0mm, clip, width = .47\linewidth]{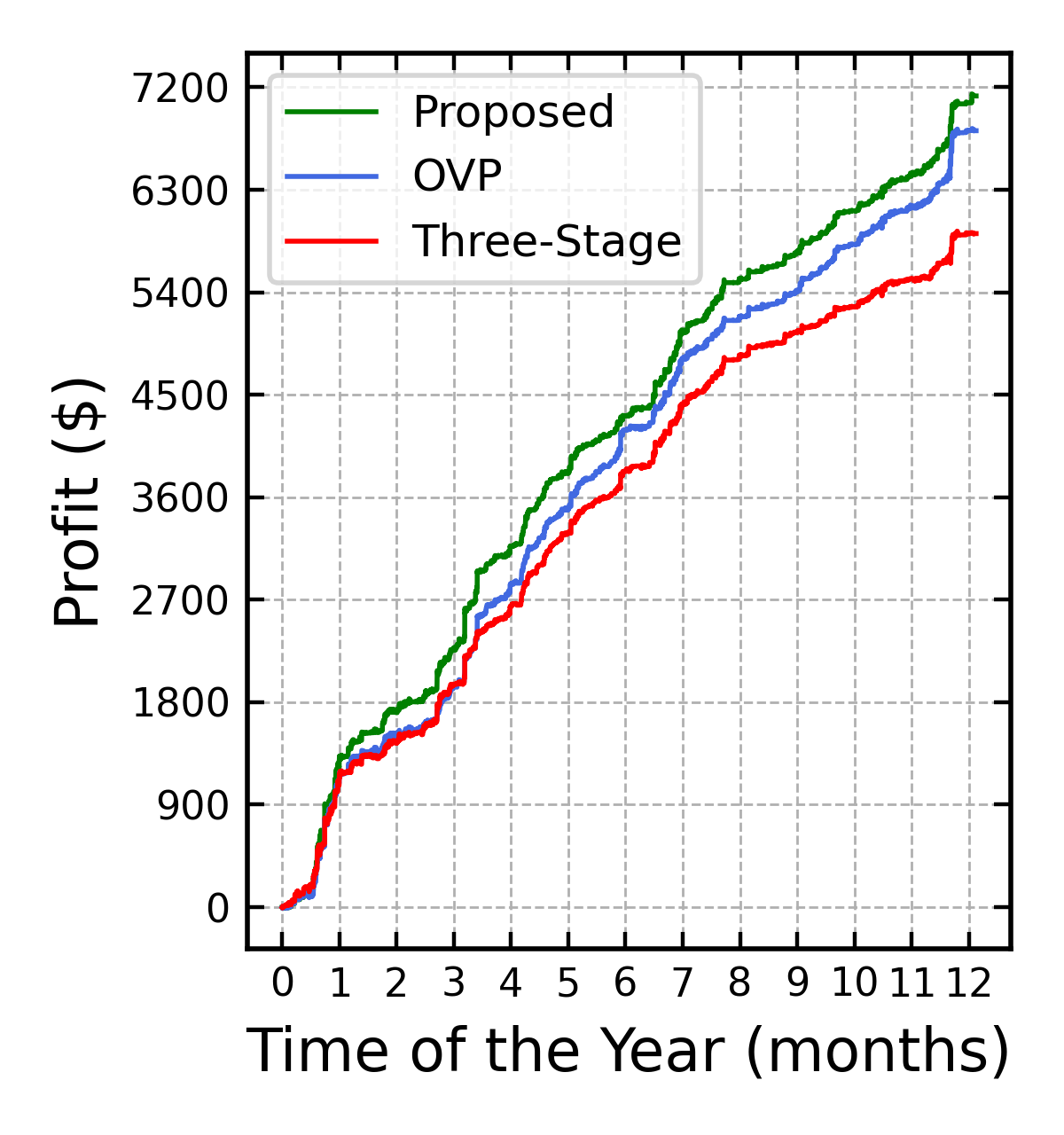}}
  \subfigure[]{ \includegraphics[trim = 3mm 0mm 3mm 0mm, clip, width = .47\linewidth]{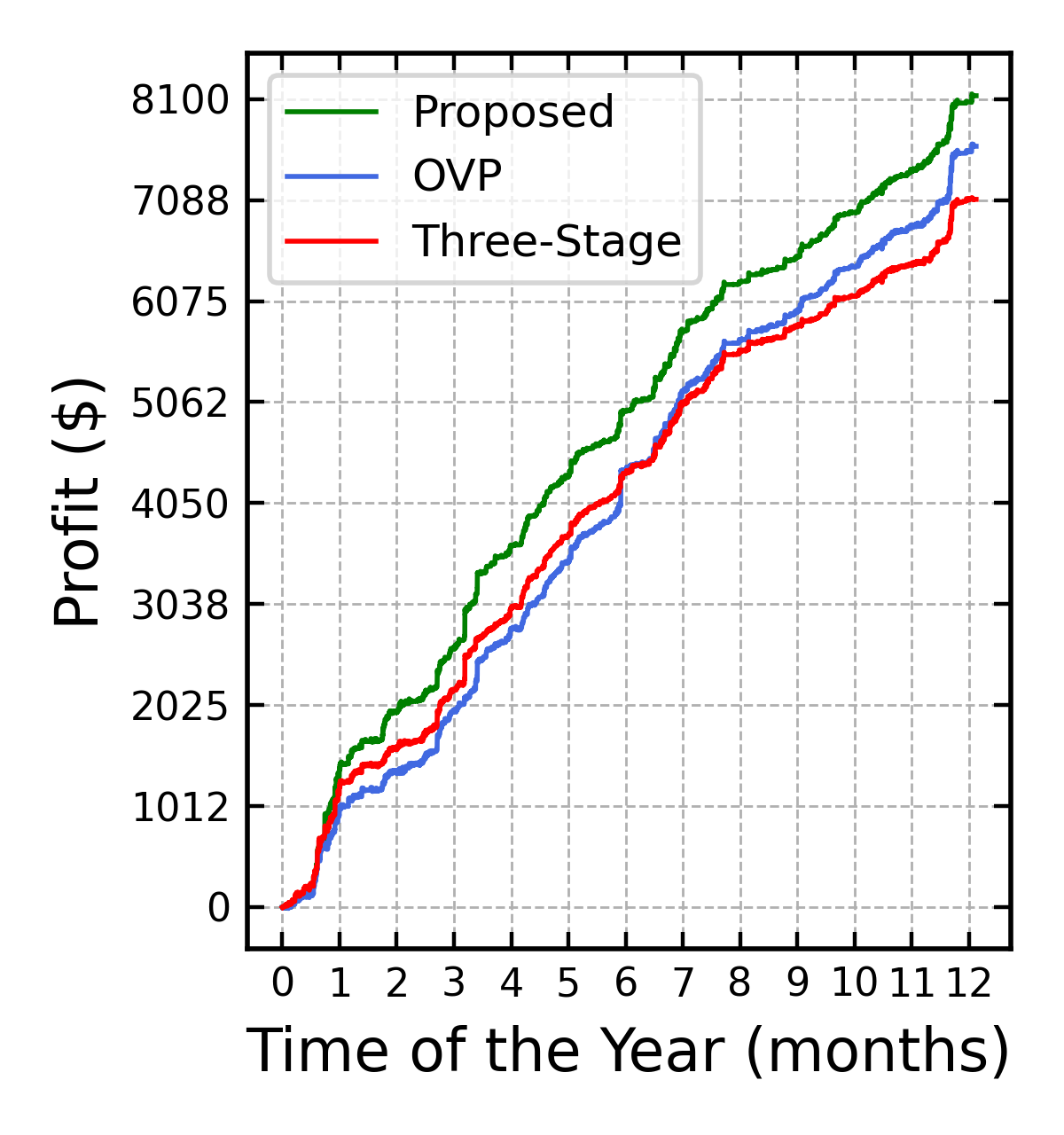} }
\vspace{-2mm}      	\caption{{Annual cumulative profit comparison for the price-taker case, 5-minute resolution real-time price data in New York City.  The energy storage model considers (a) a linear cost term, and (b) a combination of linear and quadratic cost terms.} } \label{Fig: comp_arb1_5min}
\vspace{-1em}  
\end{figure}

\begin{figure}[!ht]
  \centering
   \subfigure[]{ \includegraphics[trim = 3mm 0mm 3mm 0mm, clip, width = .48\linewidth]{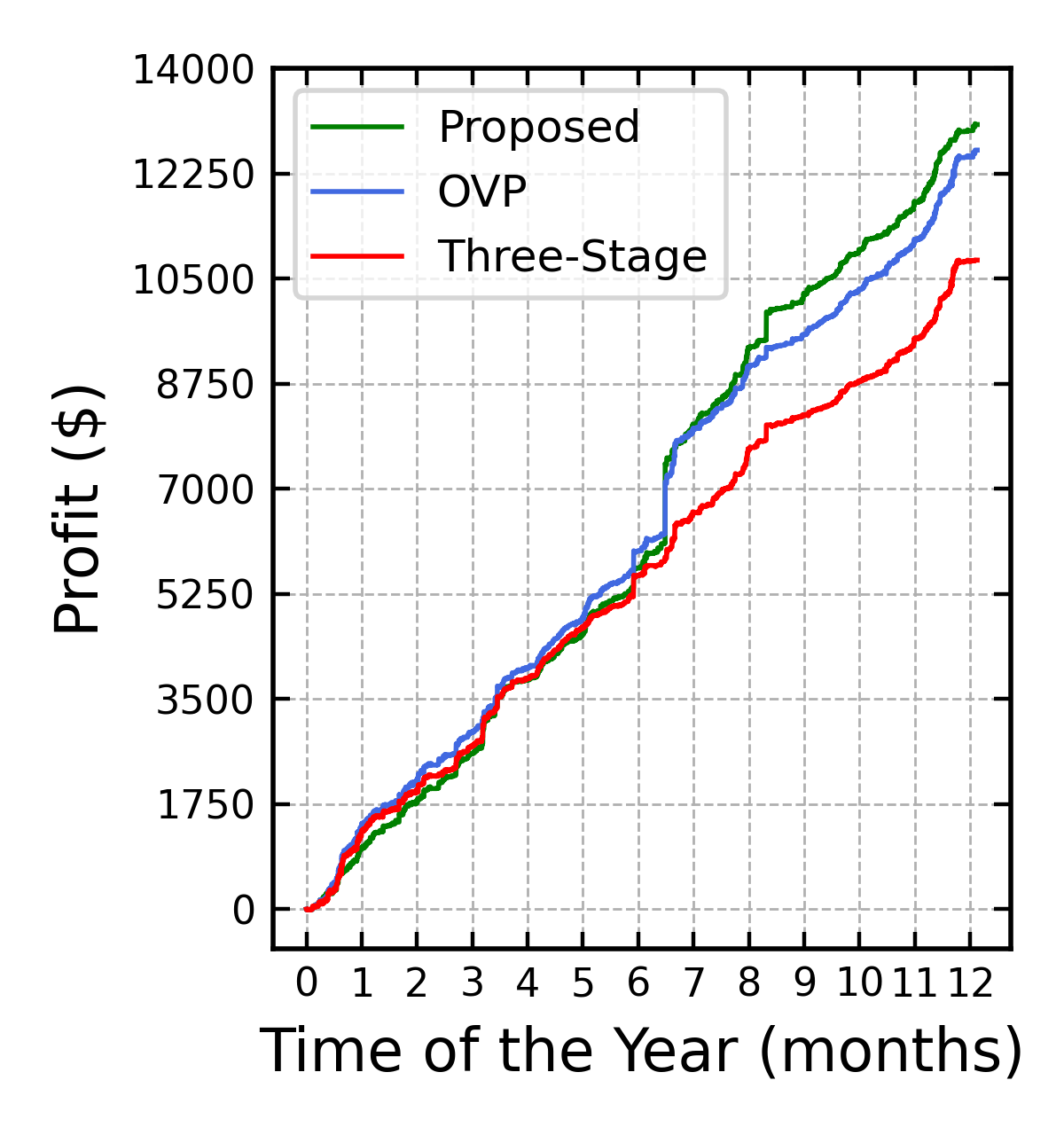}}
   \subfigure[]{ \includegraphics[trim = 3mm 0mm 3mm 0mm, clip, width = .475\linewidth]{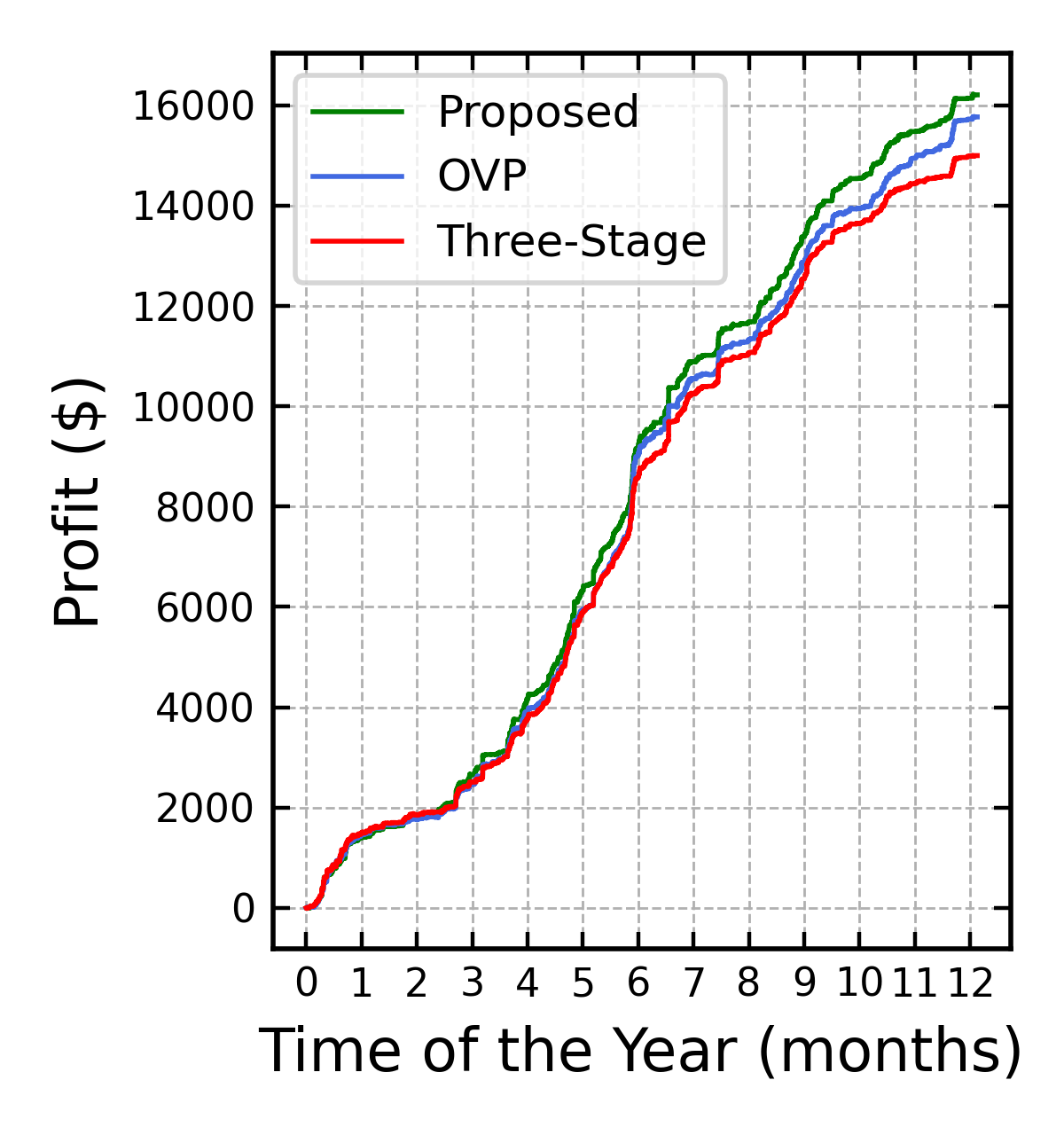}}
\vspace{-2mm}    	\caption{Annual cumulative profit comparison for the price-taker case. The energy storage model is a linear cost term for (a) Long Island and (b) Western NY in NYISO.} \label{Fig: comp_cities}

\end{figure}

% \begin{figure}[!ht]
%   \centering
%    \subfigure[]{ \includegraphics[width = .8\linewidth]{./Figures/exp1_cumsum_plot_profit.png}}
%   \subfigure[]{ \includegraphics[width = .8\linewidth]{./Figures/exp2_cumsum_plot_profit.png} }
%     	\caption{{Annual cumulative profit comparison for the price-taker case, hourly resolution real-time price data.  The energy storage model considers (a) a linear cost term, and (b) a combination of linear and quadratic cost terms.}} \label{Fig: comp_arb1_hourly}
% \end{figure}

\begin{figure}[!ht]
  \centering
   \subfigure[]{ \includegraphics[trim = 3mm 0mm 3mm 0mm, clip, width = .47\linewidth]{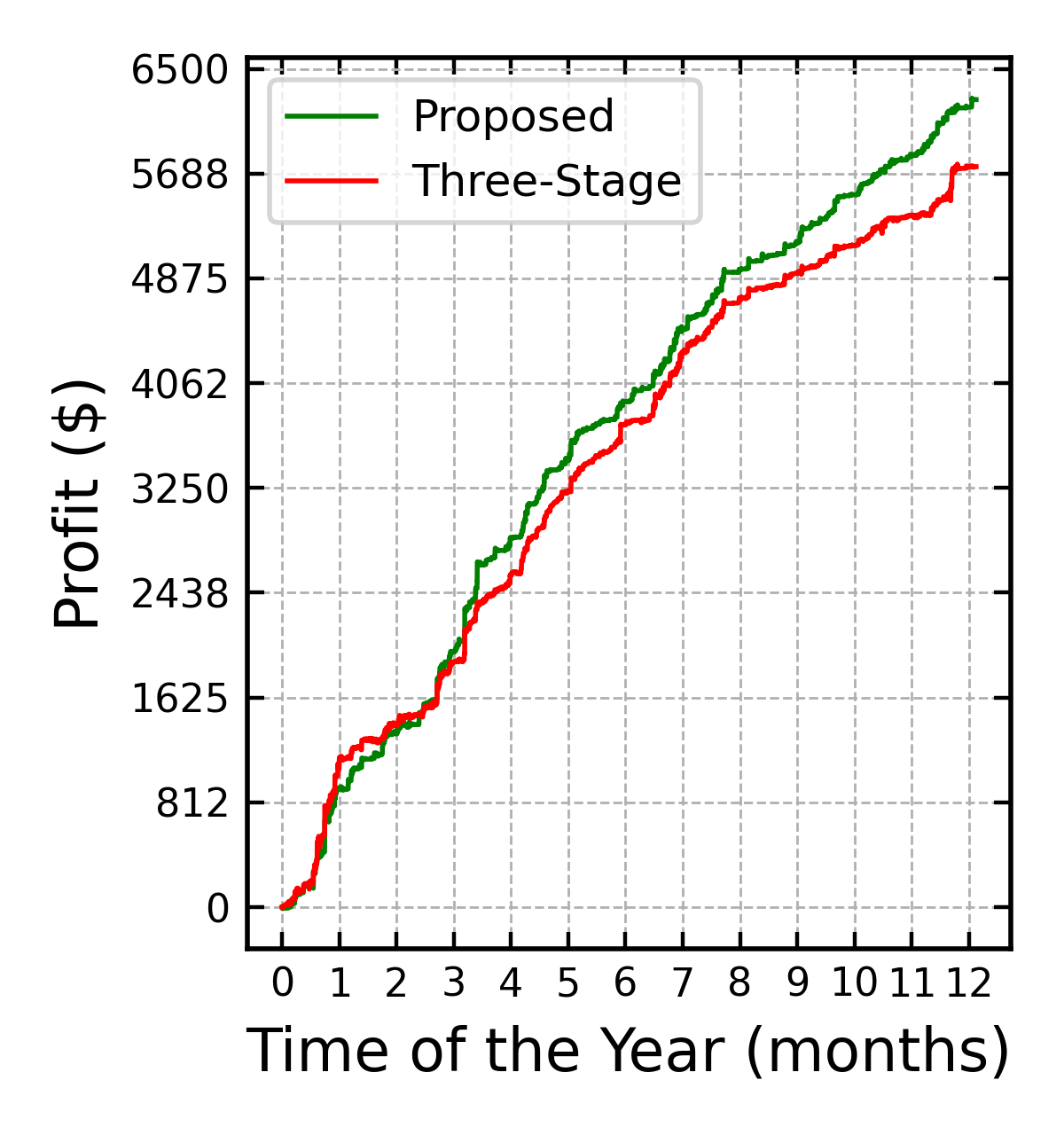}}
    \subfigure[]{ \includegraphics[trim = 3mm 0mm 3mm 0mm, clip, width = .47\linewidth]{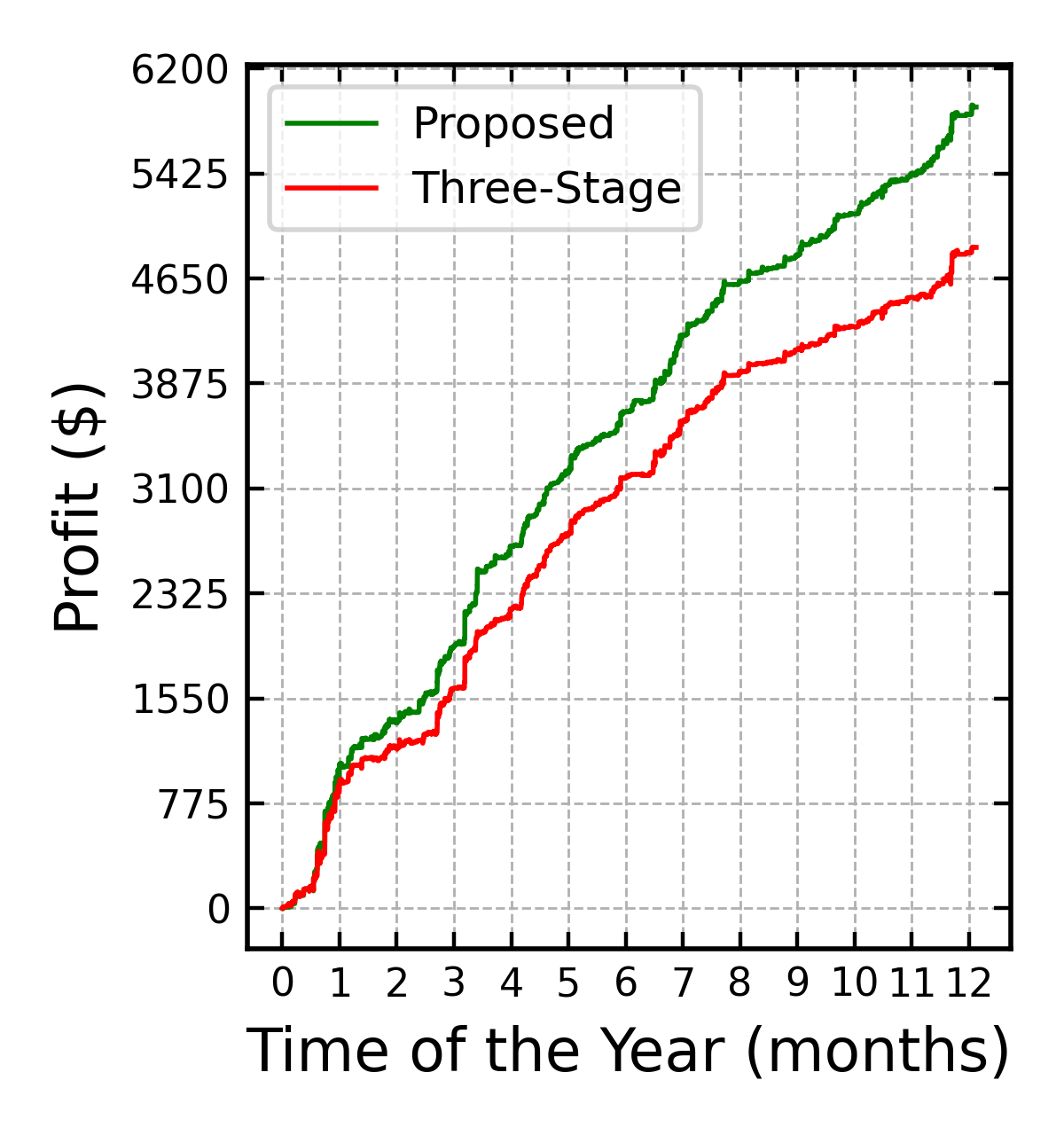}}
    \vspace{-2mm}  	\caption{Annual cumulative profit comparison for the price-maker case, considering (a) a linear price sensitivity model and (b) a cubic price sensitivity model.} \label{Fig: comp_arb2}
     \vspace{-1em}
\end{figure}

We evaluate the performance of our proposed method by comparing the cumulative profit from 2019 against two benchmark methods. The first benchmark method is called \textbf{Three-Stage}, it uses a ConvLSTM network for RTP prediction but the price prediction, bid generation, and market clearing are handled separately. The second benchmark method is called \textbf{OVP}, it uses ground-truth price data to generate the historical optimal opportunity value through dynamic programming, then trains a neural network to predict the opportunity value for the next time step and designs bids accordingly~\cite{BZX23}.
We consider an energy storage model with linear cost and a model with both linear and quadratic costs to evaluate the performance of our proposed method. 
% Note that the OVP method is based on dynamic programming and does not inherently support an energy storage model with linear and quadratic costs. For comparison purposes, we generate the opportunity value using the dual variable of the state of charge (SoC) and train a neural network to predict the opportunity value for this scenario. 
The linear cost function is defined as $c(p_t,b_t)=C_1p_t$, where $C_1= 10$. The linear and quadratic cost function is defined as $c(p_t,b_t)=C_1p_t+C_2p_t^2$, where $C_1= 5$ and $C_2=5$.  

Fig.~\ref{Fig: comp_arb1_5min} presents the cumulative profit of the proposed method compared to two benchmark methods under both energy storage models, using 5-minute resolution price data in New York City. The proposed method consistently outperforms the benchmarks, achieving a performance improvement of approximately 21\% and 5\% with the linear cost model, and 15\% and 7\% with the linear and quadratic cost model. We also compare the profit performance of the three methods in two additional NYISO zones, Long Island and Western NY, in Fig.~\ref{Fig: comp_cities}. Our method consistently achieves the highest performance. The proposed method demonstrates consistent improvements over the benchmarks.
% , with profits of approximately 32\% and 9\% under the linear cost model, and 17\% and 8\% under the linear and quadratic cost model. Our decision-focused pipeline—integrating the energy storage physical model, market clearing layer, and a decision-focused loss function—enhances overall profit performance. The results across all figures further highlight the robustness of our method under different energy storage cost functions.
\vspace{-3mm}
 \subsection{Energy Storage Bidding Under Price-Maker Setting }\label{exp:maker}

We conduct experiments on energy storage bidding under the price-maker setting. In this setting, the energy storage system can influence electricity prices by increasing prices through charging (raising demand) and decreasing prices through discharging (lowering demand). We use the same real-time price data from New York City in NYISO as described in the price-taker setting. The dataset includes both input features and optimal decision values, with optimal decisions obtained by solving \eqref{eq:clearing2} using the price-maker real-time price sensitivity model. Because the OVP method does not support the price-maker setting, we compare our method only with the Three-Stage method. Specifically, we consider two price sensitivity models: a  linear price sensitivity model, $\lambda_{t}(p_t, b_t) = {\lambda}_{t}-\alpha (p_t- b_t)$, and a cubic price sensitivity model, $\lambda_{t}(p_t, b_t) = {\lambda}_{t}-\alpha (p_t- b_t)^3$. Both models incorporate a linear degradation cost function, using the same degradation cost parameter as in the price-taker setting. 

In the testing stage, the profit is computed as follows:
\begin{equation}
    \textstyle\sum_{t=1}^T \lambda_{t}(p_t^*, b_t^*) (p_t^*-b_t^*)-c(p^*_t,b_t^*)
\end{equation}
\noindent where $p_t^*$ and $b_t^*$ denote the discharge and charge amounts, respectively, after market clearing at time $t$. The sensitivity parameter is set to $\alpha = 10$ for linear model and $\alpha = 100$ for cubic model.
 
Fig.~\ref{Fig: comp_arb2} compares the arbitrage performance of the proposed method with the Three-Stage method under price-maker settings using two price sensitivity models. The proposed method consistently outperforms the benchmark, showcasing its ability to integrate optimal decision-making under price-maker conditions. Furthermore, our end-to-end pipeline demonstrates robustness across varying sensitivity models.

%{We vary the energy storage efficiency and show the corresponding profits in Table \ref{table:arb}. The profits decrease as the efficiency decreases for all three methods. The proposed method achieves 5\%-20\% profit improvement over the two benchmark methods when $\eta=1$. In contrast, for $\eta=0.85$, the profits of the proposed method are more than double those of the two benchmark methods.
%Unlike the two benchmark methods, which treat prediction and optimization as two separate stages, the proposed method incorporates the SoC evolution into the end-to-end pipeline.  The results in Table \ref{table:arb} demonstrate that the proposed method is more robust to varying storage efficiencies, highlighting its superiority.}

%% file: Appendix.tex
\section*{Appendix}
%\setlength\columnsep{0.01in}
%\setlength\textwidth{7.08in}
% \setcounter{theorem}{0}
% \begin{proposition}  
% \emph{Marginal cost of the opportunity value function.}  Consider the predict-then-bid energy storage arbitrage problem given in \eqref{eq:1a}–\eqref{eq:1e}, assuming that both the primal and dual problems have non-empty feasible sets.  The marginal cost of the opportunity value function $V_{t+1}$  with respect to the state of charge at time 
% $t$ is defined by the dual variable of the SoC transition equation \eqref{eq:1e} at time 
% $t+1$: 
% \[
% \frac{\partial V_{t+1}}{\partial e_t}  = \bar{\theta}_{t+1}.
% \]
% The marginal costs with respect to discharge  $p_t$ and charge $b_t$ at time $t$ are  
% \begin{itemize}
%     \item  $\frac{\partial V_{t+1}}{\partial p_t} =   \frac{\partial V_{t+1}}{\partial e_t} \cdot \frac{\partial e_{t}}{\partial p_t} = -\bar{\theta}_{t+1} \cdot \frac{1}{\eta},$
%     \item  $\frac{\partial V_{t+1}}{\partial b_t} =   \frac{\partial V_{t+1}}{\partial e_t} \cdot \frac{\partial e_{t}}{\partial b_t} = \bar{\theta}_{t+1} \cdot \eta.$
% \end{itemize}  
% \end{proposition}  

\subsection{Proof of Proposition 1}

% \begin{proof}
Consider the energy storage arbitrage problem over the horizon from
$t+1$ to $T$:
\begin{align}
  \underset{p_k, b_k \in \mathcal{X}_1}{\textbf{max}}  \quad \sum_{k=t+1}^{T}\left(\hat{\lambda}_k \left( p_k - b_k \right) - c(p_k,b_k)\right).
\end{align}

\noindent The Lagrangian function of this problem is given by:
\begin{align}
    \mathbb{L}(p&,b,e)  \nonumber\\
    = &  \sum_{k={t+1}}^T \left( \hat{\lambda}_k (p_k - b_k) - c(p_k,b_k) \right) \nonumber \\
    &- \sum_{k={t+1}}^T \bar{\theta}_k \left( e_k - e_{k-1} +  \frac{p_k}{\eta} - b_k \eta \right)  \nonumber \\&- \sum_{k={t+1}}^T \left( \bar{u}_k( p_k - P) - \underline{u}_k p_k + \bar{v}_k( b_k - P) - \underline{v}_k b_k \right) \nonumber \\ &  - \sum_{k={t+1}}^T \left( \bar{w}_k( e_k - E) - \underline{w}_k e_k \right). \label{eq:lagrangian}
\end{align}

Let $p_k^*, b_k^*, e_k^*, \bar{\theta}_{k}^*, \bar{u}_k^*, \underline{u}_k^*, \bar{w}_k^*, \underline{w}_k^*$ denote the optimal solutions that maximize the objective function under constraints. Substituting these values into \eqref{eq:lagrangian} to obtain  $\mathbb{L}^*(p,b,e)$.  The value function is then defined as:
\begin{align}
    V_{t+1} = &  \sum_{k={t+1}}^T \left( \hat{\lambda}_k (p_k^* - b_k^*) - c(p_k^*,b_k^*) \right) = \mathbb{L}^*(p,b,e)
\end{align}
The equality $V_{t+1} = \mathbb{L}^*(p,b,e)$ holds because the inequality constraints \eqref{eq:1b}-\eqref{eq:1d} satisfy complementary slackness.  To compute the marginal cost of the opportunity value with respect to $e_t$, note that changes in $b_t$  and $p_t$ influence $e_t$, thereby affecting future energy levels. According to the envelope theorem \cite{C01}, at the optimum, only the direct effect of $e_t$ on the objective function needs to be considered. Thus, the partial derivative of  $V_{t+1}$ with respect to $e_t$  is given by:
%\begin{align}
%    \frac{\partial V_{t+1}}{\partial e_t} = \frac{\partial \mathbb{L}^*(p,b,e)}{\partial e_t} 
%\end{align}
\begin{align}
 \frac{\partial V_{t+1}}{\partial e_t} =& \frac{\partial \mathbb{L}^*(p,b,e)}{\partial e_t} \\
 =&   \frac{\partial}{\partial e_t} \Bigg( - \sum_{k={t+1}}^T \bar{\theta}_k \left( e_k - e_{k-1} + \frac{p_k}{\eta} - b_k \eta \right) \nonumber
 \\&- \sum_{k={t+1}}^T \left( \bar{w}_k( e_k - E) - \underline{w}_k e_k \right) \Bigg) = \bar{\theta}_{t+1}
\end{align}

Because charge and discharge influence  the state of charge through efficiency, the marginal cost of the opportunity value with respect to $p_t$ and $b_t$ is derived using the chain rule:
\begin{subequations}
\begin{align}\label{dual1}
   \frac{\partial V_{t+1}}{\partial p_t} &=   \frac{\partial V_{t+1}}{\partial e_t}    \frac{\partial e_{t}}{\partial p_t}=-\frac{\partial V_{t+1}}{\partial e_t}\cdot \frac{1}{\eta} = -\bar{\theta}_{t+1}  \cdot\frac{1}{\eta} \\
\label{dual2} 
   \frac{\partial V_{t+1}}{\partial b_t} &=   \frac{\partial V_{t+1}}{\partial e_t}    \frac{\partial e_{t}}{\partial b_t}=\frac{\partial V_{t+1}}{\partial e_t}\cdot \eta =\bar{\theta}_{t+1} \cdot \eta
\end{align}
\end{subequations}

\noindent where $\bar{\theta}_{t+1}$ denotes the dual variable corresponding to the SoC dynamic constraint in equation \eqref{eq:1e} at time $t+1$.

% \end{proof}

% \setcounter{theorem}{1}
% \begin{proposition}  
% \emph{Differentiation of the dual variable of SoC transition.}  Consider the Lagrangian associated with the optimization problem \eqref{eq:1a}–\eqref{eq:1e}. Let \(\bm{\mu}\) denote the dual variables corresponding to all inequality constraints. The Karush-Kuhn-Tucker (KKT) conditions define the function:
% \[
% \mathcal{F}(\bm{p}, \bm{b}, \bm{e}, \hat{\bm{\lambda}}, \bar{\bm{\theta}}_t, \bm{\mu}) = 0.
% \]

% \noindent Based on the Implicit Function Theorem and chain rule, differentiating the KKT system with respect to \(\hat{\bm{\lambda}}\) gives:
% \[
% \frac{\partial \mathcal{F}}{\partial \hat{\bm{\lambda}}} 
% + \frac{\partial \mathcal{F}}{\partial \bar{\bm{\theta}}_t} 
% \frac{\partial \bar{\bm{\theta}}_t }{\partial \hat{\bm{\lambda}}} = 0.
% \]

% \noindent Rearrange this equation to obtain the derivative:
% \[
% \frac{\partial \bar{\bm{\theta}}_t }{\partial \hat{\bm{\lambda}}} = 
% - \left( \frac{\partial \mathcal{F}}{\partial \bar{\bm{\theta}}_t} \right)^{-1} 
% \frac{\partial \mathcal{F}}{\partial \hat{\bm{\lambda}}}.
% \]
% \end{proposition}  

\subsection{Proof of Proposition 2}
% \begin{proof}

Given the storage arbitrage model~\eqref{eq:1a}–\eqref{eq:1e}, the Lagrangian function can be expressed as:
\begin{subequations}
    \begin{align}
        L 
        % = &  \sum_{k={t}}^T \left( \hat{\lambda}_k (p_k - b_k) - c(p_k,b_k) \right) \nonumber \\
        % &- \sum_{k={t}}^T \bar{\theta}_k \left( e_k - e_{k-1} +  \frac{p_k}{\eta} - b_k \eta \right)  \nonumber \\
        % &- \sum_{k={t}}^T \left( \bar{u}_k( p_k - R) - \underline{u}_k p_k + \bar{v}_k( b_k - R) - \underline{v}_k b_k \right) \nonumber \\ 
        % &  - \sum_{k={t}}^T \left( \bar{w}_k( e_k - E) - \underline{w}_k e_k \right) \\
        = &  \sum_{k={t}}^T \left( \hat{\lambda}_k (p_k - b_k) - c(p_k,b_k) \right) - \sum_{k={t}}^T \bar{\theta}_k \left(  \frac{p_k}{\eta} - b_k \eta \right)  \nonumber \\
        &- \sum_{k={t}}^T \left( \bar{u}_k( p_k - R) - \underline{u}_k p_k + \bar{v}_k( b_k - R) - \underline{v}_k b_k \right) \nonumber \\ 
        &  + \sum_{k={t}}^T  \bar{w}_k  E    + \sum_{k={t}}^{T-1} \left(-\bar{\theta}_k+\bar{\theta}_{k+1}-\bar{w}_k+\underline{w}_k\right)e_k \nonumber \\
        & +(-{\theta}_T-\bar{w}_T+\underline{w}_T)e_T + {\theta}_t e_{t-1}
    \end{align}
    \label{eq:offline_lag}
\end{subequations}

\noindent The KKT conditions $\mathcal{F}(\bm{p}, \bm{b}, \bm{e}, \hat{\bm{\lambda}}, \bar{\bm{\theta}}_t, \bm{\mu})$ can be written as follows:
\begin{subequations}
    \begin{align} 
        & \nabla_{p_t} L =\hat{\lambda}_{t}  - \frac{\partial c(p_t,b_t) }{\partial p_t} -\frac{\bar{\theta}_t}{\eta} -  \bar{u}_t + \underline{u}_t = 0, \label{KKT_dual_1}\\
        & \nabla_{b_t} L = -\hat{\lambda}_{t} - \frac{\partial c(p_t,b_t) }{\partial b_t}  + \bar{\theta}_t \eta - \bar{v}_t + \underline{v}_t = 0, \label{KKT_dual_2}\\
        & \nabla_{e_t} L = -\bar{\theta}_t + \bar{\theta}_{t+1} - \bar{w}_t + \underline{w}_t = 0, \label{KKT_dual_3} \\
        & \nabla_{e_T} L = -\bar{\theta}_T - \bar{w}_T + \underline{w}_T = 0, \label{KKT_dual_4}\\
        & e_{t} - e_{t-1} = -\frac{p_{t}}{\eta} + b_{t} \eta,   \quad   0 \leq e_{t} \leq E, \label{KKT_primal_1}\\
        & 0 \leq p_{t} \leq R, \quad 0 \leq b_{t} \leq R, \label{KKT_primal_2}\\
        & \bar{u}_t, \underline{u}_t, \bar{v}_t, \underline{v}_t, \bar{w}_t, \underline{w}_t \geq 0, \label{KKT_dual_5} \\
        &  \bar{u}_t (R - p_{t}) = 0, \quad \underline{u}_t (p_{t}) = 0, \label{KKT_comp_1}\\
        &  \bar{v}_t (R - b_{t}) = 0, \quad \underline{v}_t (b_{t}) = 0,\label{KKT_comp_2} \\
        & \bar{w}_t (E - e_{t}) = 0, \quad \underline{w}_t (e_{t}) = 0 \label{KKT_comp_3}
    \end{align}
\end{subequations}
where equations~\eqref{KKT_dual_1}–\eqref{KKT_dual_4} denote the stationarity conditions of the arbitrage problem~\eqref{eq:1a}–\eqref{eq:1e}; \eqref{KKT_primal_1}–\eqref{KKT_primal_2} specify the primal feasibility conditions; \eqref{KKT_dual_5} denotes the dual feasibility condition; and \eqref{KKT_comp_1}–\eqref{KKT_comp_3} represent the complementary slackness conditions.

\begin{figure*}[t]
    \begin{footnotesize}
        \begin{equation}
            \underbrace{
\begin{pmatrix}
\pmb{0} & \pmb{0} & \pmb{0} & -\frac{1}{\eta} \pmb{I} & -\pmb{I} & \pmb{I} & \pmb{0} & \pmb{0} & \pmb{0} & \pmb{0} \\
\pmb{0} & \pmb{0} & \pmb{0} & \eta \pmb{I} & \pmb{0} & \pmb{0} & -\pmb{I} & \pmb{I} & \pmb{0} & \pmb{0} \\
\pmb{0} & \pmb{0} & \pmb{0} & (\bm{\Gamma} - \pmb{I}) & \pmb{0} & \pmb{0} & \pmb{0} & \pmb{0} & -\pmb{I} & \pmb{I} \\
\frac{1}{\eta} \pmb{I} & -\eta \pmb{I} & (\pmb{I} -\bm{\Gamma}^\top  ) & \pmb{0} & \pmb{0} & \pmb{0} & \pmb{0} & \pmb{0} & \pmb{0} & \pmb{0} \\
-\Lambda(\bar{\pmb{u}}) & \pmb{0} & \pmb{0} & \pmb{0} & \Lambda(\mathds{1}R - \pmb{p}) & \pmb{0} & \pmb{0} & \pmb{0} & \pmb{0} & \pmb{0} \\
\Lambda(\underline{\pmb{u}}) & \pmb{0} & \pmb{0} & \pmb{0} & \pmb{0} & \Lambda(\pmb{p}) & \pmb{0} & \pmb{0} & \pmb{0} & \pmb{0} \\
\pmb{0} & -\Lambda(\bar{\pmb{v}}) & \pmb{0} & \pmb{0} & \pmb{0} & \pmb{0} & \Lambda(\mathds{1}R - \pmb{b}) & \pmb{0} & \pmb{0} & \pmb{0} \\
\pmb{0} & \Lambda(\underline{\pmb{v}}) & \pmb{0} & \pmb{0} & \pmb{0} & \pmb{0} & \pmb{0} & \Lambda(\pmb{b}) & \pmb{0} & \pmb{0} \\
\pmb{0} & \pmb{0} & -\Lambda(\bar{\pmb{w}}) & \pmb{0} & \pmb{0} & \pmb{0} & \pmb{0} & \pmb{0} & \Lambda(\mathds{1}E - \pmb{e}) & \pmb{0} \\
\pmb{0} & \pmb{0} & \Lambda(\underline{\pmb{w}}) & \pmb{0} & \pmb{0} & \pmb{0} & \pmb{0} & \pmb{0} & \pmb{0} & \Lambda(\pmb{e})
\end{pmatrix}}_{\bm{A}}
\begin{pmatrix}
\mathrm{d}\pmb{p} \\ \mathrm{d}\pmb{b} \\ \mathrm{d}\pmb{e} \\ \mathrm{d}\bar{\pmb{\theta}}_t \\ \mathrm{d}\bar{\pmb{u}} \\ \mathrm{d}\underline{\pmb{u}} \\ \mathrm{d}\bar{\pmb{v}} \\ \mathrm{d}\underline{\pmb{v}} \\ \mathrm{d}\bar{\pmb{w}} \\ \mathrm{d}\underline{\pmb{w}}
\end{pmatrix}
=
\begin{pmatrix}
-\mathrm{d}\hat{\pmb{\lambda}} \\
\mathrm{d}\hat{\pmb{\lambda}} \\
\pmb{0} \\
\pmb{0} \\
\pmb{0} \\
\pmb{0} \\
\pmb{0} \\
\pmb{0} \\
\pmb{0}
\end{pmatrix}
\label{eq:total_dev}
\end{equation}
\end{footnotesize}
\end{figure*}

The total derivative of the KKT conditions~\eqref{KKT_dual_1}–\eqref{KKT_comp_3} is taken as follows:
\begin{subequations}
    \begin{align} 
        &  \mathrm{d}\hat{\pmb{\lambda}}   -\frac{\mathrm{d}\bar{\pmb{\theta}}_t}{\eta} - \mathrm{d}\bar{\pmb{u}} + \mathrm{d}\underline{\pmb{u}} = \pmb{0}, \label{KKT_dual_11}\\
        &  -\mathrm{d}\hat{\pmb{\lambda}}  + \mathrm{d}\bar{\pmb{\theta}}_t \eta - \mathrm{d}\bar{\pmb{v}} + \mathrm{d}\underline{\pmb{v}} = \pmb{0}, \label{KKT_dual_22}\\
        & ( \bm{\Gamma}-\pmb{I})\mathrm{d}\bar{\pmb{\theta}}_t  - \mathrm{d}\bar{\pmb{w}} + \mathrm{d}\underline{\pmb{w}} = \pmb{0}, \label{KKT_dual_33 } \\
       % & \nabla_{e_T} L = -d\theta_T - d\bar{w}_T + d\underline{w}_T = \pmb{0}, \label{KKT_dual_4}\\
     &  (\pmb{I} - \bm{\Gamma}^\top) \mathrm{d}\pmb{e} +\frac{1}{\eta} \mathrm{d}\pmb{p} - \eta \mathrm{d}\pmb{b}=0, \label{KKT_dual_44}\\
        & \Lambda (\mathds{1} R - \pmb{p}) \mathrm{d}\bar{\pmb{u}}+ \Lambda (  \bar{\pmb{u}} )( - \mathrm{d}\pmb{p}) = \pmb{0}, \quad \Lambda (\pmb{p} ) \mathrm{d}\underline{\pmb{u}} +\Lambda ( \underline{\pmb{u}} ) \mathrm{d}\pmb{p} = \pmb{0}, \label{KKT_comp_11}\\
        & \Lambda (\mathds{1}R - \pmb{b}  )\mathrm{d} \bar{\pmb{v}}+ \Lambda ( \bar{\pmb{v}} )(- \mathrm{d}\pmb{b}) = \pmb{0}, \quad  \Lambda (\pmb{b}) \mathrm{d}\underline{\pmb{v} }+\Lambda (\underline{\pmb{v}}) \mathrm{d}\pmb{b} = \pmb{0},\label{KKT_comp_22} \\
        &\Lambda (\mathds{1}E - \pmb{e})\mathrm{d}\bar{\pmb{w}}+ \Lambda ( \bar{\pmb{w}}) ( - \mathrm{d}\pmb{e}) = \pmb{0}, \quad \Lambda (\pmb{e})\mathrm{d}\underline{\pmb{w}} +\Lambda (\underline{\pmb{w}})\mathrm{d}\pmb{e} = \pmb{0} \label{KKT_comp_33}
    \end{align}
\end{subequations}
%$\bar{\pmb{\theta}}_t=[\bar{\theta}_t, \bar{\theta}_{t+1},...,\bar{\theta}_T]$ and 
%$\bar{\pmb{\theta}}_{t+1}=[\bar{\theta}_{t+1}, \bar{\theta}_{t+2},..., \bar{\theta}_{T}, 0]$. The notation $\Lambda(.)$ represents the operation of creating a diagonal matrix from a given vector. $\bm{A}$ denotes the left-hand side coefficient matrix in~\eqref{eq:total_dev}. 

\noindent where $\mathds{1} \in \mathbb{R}^T$ denotes an all-one vector. $\bar{\pmb{\theta}}_t=[\bar{\theta}_t, \bar{\theta}_{t+1},...,\bar{\theta}_T]$ and 
$\bar{\pmb{\theta}}_{t+1}=[\bar{\theta}_{t+1}, \bar{\theta}_{t+2},..., \bar{\theta}_{T}, 0]$. The notation $\Lambda(.)$ represents the operation of creating a diagonal matrix from a given vector.  $\bm{\Gamma}$ is a shift matrix to link the state at one-time step to the next state in \eqref{KKT_dual_3} and \eqref{KKT_dual_4}.  Then \( \mathrm{d}\pmb{\theta}_{t+1} = \bm{\Gamma} \mathrm{d}\bar{\pmb{\theta}}_t \), and the difference \( \mathrm{d}\pmb{\theta}_{t+1} - \mathrm{d}\bar{\pmb{\theta}}_t \) becomes $\bm{\Gamma} \mathrm{d}\bar{\pmb{\theta}}_t - \mathrm{d}\bar{\pmb{\theta}}_t$. The matrix $\bm{\Gamma}$ is defined as:
\[
\bm{\Gamma} = \begin{pmatrix}
0 & 1 & 0 & \cdots & 0 \\
0 & 0 & 1 & \cdots & 0 \\
0 & 0 & 0 & \ddots & 0 \\
\vdots & \vdots & \vdots & \ddots & 1 \\
0 & 0 & 0 & \cdots & 0
\end{pmatrix}
\]

The equations~\eqref{KKT_dual_11}--\eqref{KKT_comp_33} can be written in a compact matrix form as in~\eqref{eq:total_dev}, where we denote $\bm{A}$ as the left-hand side coefficient matrix in~\eqref{eq:total_dev}. Both sides of~\eqref{eq:total_dev} are divided by \(\mathrm{d} \hat{\pmb{\lambda}}\) and multiplied by \(\bm{A}^{-1}\) to obtain the gradient with respect to the price prediction:

\begin{align}&
\begin{pmatrix}
\frac{\mathrm{d}\pmb{p}}{\mathrm{d} \hat{\pmb{\lambda}}},  \frac{\mathrm{d}\pmb{b}}{\mathrm{d} \hat{\pmb{\lambda}}},  \frac{\mathrm{d}\pmb{e}}{\mathrm{d} \hat{\pmb{\lambda}}},  \frac{\mathrm{d}\bar{\pmb{\theta}}_t}{\mathrm{d} \hat{\pmb{\lambda}}},  \frac{\mathrm{d}\bar{\pmb{u}}}{\mathrm{d} \hat{\pmb{\lambda}}},  \frac{\mathrm{d}\underline{\pmb{u}}}{\mathrm{d} \hat{\pmb{\lambda}}},  \frac{\mathrm{d}\bar{\pmb{v}}}{\mathrm{d} \hat{\pmb{\lambda}}},  \frac{\mathrm{d}\underline{\pmb{v}}}{\mathrm{d} \hat{\pmb{\lambda}}},  \frac{\mathrm{d}\bar{\pmb{w}}}{\mathrm{d} \hat{\pmb{\lambda}}},  \frac{\mathrm{d}\underline{\pmb{w}}}{\mathrm{d} \hat{\pmb{\lambda}}} 
\end{pmatrix}^{\mathsf{T}}  \nonumber \\
& \qquad\qquad\qquad = \bm{A}^{-1} \begin{pmatrix}
-\mathds{1}  ,
\mathds{1} , 
\pmb{0} ,
\pmb{0}, 
\pmb{0} ,
\pmb{0},
\pmb{0} ,
\pmb{0} ,
\pmb{0}
\end{pmatrix}^{\mathsf{T}}\label{eq:grad}
\end{align}

%Next, we can divide both sides of~\eqref{eq:total_dev} with the input derivative 
% $
% \begin{pmatrix}
% d\pmb{p},  d\pmb{b},  d\pmb{e},  d\bar{\pmb{\theta}}_t,  d\bar{\pmb{u}},  d\underline{\pmb{u}},  d\bar{\pmb{v}},  d\underline{\pmb{v}},  d\bar{\pmb{w}},  d\underline{\pmb{w}}
% \end{pmatrix}$  by 
%$\mathrm{d} \hat{\pmb{\lambda}}$ and multiply both sides with $\bm{A}^{-1}$ to infer the gradient with respect to the cost term and price prediction:
%\begin{align}&
%\begin{pmatrix}
% \frac{\mathrm{d}\pmb{p}}{\mathrm{d} \hat{\pmb{\lambda}}},  \frac{\mathrm{d}\pmb{b}}{\mathrm{d} \hat{\pmb{\lambda}}},  \frac{\mathrm{d}\pmb{e}}{\mathrm{d} \hat{\pmb{\lambda}}},  \frac{\mathrm{d}\bar{\pmb{\theta}}_t}{\mathrm{d} \hat{\pmb{\lambda}}},  \frac{\mathrm{d}\bar{\pmb{u}}}{\mathrm{d} \hat{\pmb{\lambda}}},  \frac{\mathrm{d}\underline{\pmb{u}}}{\mathrm{d} \hat{\pmb{\lambda}}},  \frac{\mathrm{d}\bar{\pmb{v}}}{\mathrm{d} \hat{\pmb{\lambda}}},  \frac{\mathrm{d}\underline{\pmb{v}}}{\mathrm{d} \hat{\pmb{\lambda}}},  \frac{\mathrm{d}\bar{\pmb{w}}}{\mathrm{d} \hat{\pmb{\lambda}}},  \frac{\mathrm{d}\underline{\pmb{w}}}{\mathrm{d} \hat{\pmb{\lambda}}} 
% \end{pmatrix}^{\mathsf{T}}  \nonumber \\
% & \qquad\qquad\qquad = \bm{A}^{-1} \begin{pmatrix}
% -\mathds{1}  ,
% \mathds{1} , 
% \pmb{0} ,
% \pmb{0}, 
% \pmb{0} ,
% \pmb{0},
% \pmb{0} ,
% \pmb{0} ,
% \pmb{0}
% \end{pmatrix}^{\mathsf{T}}\label{eq:grad}
% \end{align}
\noindent The matrix \(\frac{\partial \bar{\pmb{\theta}}_t}{\partial \hat{\pmb{\lambda}}} \in \mathbb{R}^{T \times T}\) is obtained as the fourth component in~\eqref{eq:grad}. We extract its first row and repeat this process for each of the \(N\) initial SoC levels. The resulting gradients from all segments are then stacked to construct the final Jacobian matrix \(\frac{\partial \pmb{\theta}}{\partial \hat{\pmb{\lambda}}} \in \mathbb{R}^{N \times T}\).
% \end{proof}

% \begin{proposition} 
% \emph{Differentiability of loss function with perturbed bid.}  Assume the noise $\bm{Z}$ follows a Gaussian distribution with density $\vartheta(\bm{Z}) \propto \exp(-\psi(\bm{Z}))$. For notational simplicity, define $\bm{y}_{\epsilon}^* = \bm{p}_\epsilon^*-\bm{b}_\epsilon^*$.
%  \begin{itemize}
%  \item The perturbed loss function $\mathcal{L}^{DF}_\epsilon$ is twice differentiable, and 
% $\nabla_{\bar{\bm{\theta}}_{t+1}} F_{\epsilon}(\bar{\bm{\theta}}_{t+1})=\bm{y}_{\epsilon}^*(\bar{\bm{\theta}}_{t+1}) =\mathbb{E}[ y^*(\bar{\bm{\theta}}_{t+1}+\epsilon \bm{Z})]=\mathbb{E}[F(\bar{\bm{\theta}}_{t+1}+\epsilon \bm{Z})\nabla_Z \psi(\bm{Z})/\epsilon]$. 
% \item $\bm{y}_{\epsilon}^*$ is differentiable, and  $J_{\theta}\bm{y}_{\epsilon}^*=\mathbb{E}[F_2(\bar{\bm{\theta}}_{t+1}+\epsilon \bm{Z})(\vartheta(\bm{Z})\vartheta^{\top}(\bm{Z})-\nabla^2\vartheta(\bm{Z}) ) /\epsilon]$
% \end{itemize}  

% \end{proposition}  

\subsection{Proof of Proposition 3}
% \begin{proof}

The key to proving that the perturbed loss function is differentiable with respect to the bid is to show differentiability in terms of $F_{2,\epsilon}(\bar{\bm{\theta}}_{t+1})$.

\noindent We define:
\begin{equation}
    \bm{G} = \bar{\bm{\theta}}_{t+1} + \epsilon \bm{Z},
\end{equation}
which leads to:
\begin{equation}
    \bm{Z} = \frac{\bm{G} - \bar{\bm{\theta}}_{t+1}}{\epsilon}
\end{equation}
Differentiate both sides, we obtain:
\begin{equation}
    d\bm{Z} = \frac{d\bm{G}}{\epsilon}
\end{equation}

Use the definition of \( F_{\epsilon}(\bar{\bm{\theta}}_{t+1}) \) and apply the change of variables, we have:
\begin{equation} \label{proof: prop1}
\begin{aligned}
    F_{2, \epsilon}(\bar{\bm{\theta}}_{t+1}) &= \mathbb{E}[ F_2(\bar{\bm{\theta}}_{t+1} + \epsilon \bm{Z})] \\
    &= \int F_2(\bar{\bm{\theta}}_{t+1} + \epsilon \bm{Z}) \vartheta(\bm{Z}) d\bm{Z}  \\
    &= \frac{1}{\epsilon} \int F_2(\bm{G}) \vartheta\left(\frac{\bm{G} - \bar{\bm{\theta}}_{t+1}}{\epsilon}\right) d\bm{G}.
\end{aligned}
\end{equation}

Since the noise \( \bm{Z} \) follows a Gaussian distribution, \( \psi(\bm{Z}) \) is differentiable, and we have:
\begin{equation}
    \nabla \vartheta(\bm{Z}) = - \vartheta(\bm{Z}) \nabla \psi(\bm{Z}).
\end{equation}

Differentiate with respect to \( \bar{\bm{\theta}}_{t+1} \) and reverse the change of variables, we obtain:
\begin{equation} \label{eq: prop2}
\begin{aligned}
   & \nabla_{\bar{\bm{\theta}}_{t+1}} F_{2,\epsilon}(\bar{\bm{\theta}}_{t+1}) \\ &= \frac{1}{\epsilon} \int F_2(\bm{G}) \nabla_{\bar{\bm{\theta}}_{t+1}} \vartheta\left(\frac{\bm{G} - \bar{\bm{\theta}}_{t+1}}{\epsilon}\right) d\bm{G}  \\
    &= \frac{1}{\epsilon^2} \int F_2(\bm{G}) \vartheta \left(\frac{\bm{G} - \bar{\bm{\theta}}_{t+1}}{\epsilon}\right) \nabla_{\bar{\bm{\theta}}_{t+1}} \psi\left(\frac{\bm{G} - \bar{\bm{\theta}}_{t+1}}{\epsilon}\right) d\bm{G}  \\
    &= \frac{1}{\epsilon} \int F_2(\bar{\bm{\theta}}_{t+1} + \epsilon \bm{Z}) \vartheta(\bm{Z}) \nabla_{\bm{Z}} \psi(\bm{Z}) d\bm{Z}  \\
    &= \mathbb{E} \left[ F_2(\bar{\bm{\theta}}_{t+1} + \epsilon \bm{Z}) \nabla_{\bm{Z}} \psi(\bm{Z}) / \epsilon \right].
\end{aligned}
\end{equation}

If $\psi(\bm{Z})$ is twice differentiable, then:
\begin{equation}
    \nabla^2 \vartheta(\bm{Z}) = (\vartheta(\bm{Z}) \vartheta^{\top}(\bm{Z}) - \nabla^2 \vartheta(\bm{Z}) ) \vartheta(\bm{Z}).
\end{equation}

Thus, differentiating \( \nabla_{\bar{\bm{\theta}}_{t+1}} F_{2,\epsilon}(\bar{\bm{\theta}}_{t+1}) \) once more:
\begin{equation} \label{eq: prop3}
\begin{aligned}
    &\nabla^2_{\bar{\bm{\theta}}_{t+1}} F_{2,\epsilon}(\bar{\bm{\theta}}_{t+1}) \\&= \frac{1}{\epsilon} \int F_2(\bm{G}) \nabla^2_{\bar{\bm{\theta}}_{t+1}} \vartheta^2 \left(\frac{\bm{G} - \bar{\bm{\theta}}_{t+1}}{\epsilon}\right) d\bm{G}  \\
    &= \frac{1}{\epsilon} \int F_2(\bar{\bm{\theta}}_{t+1} + \epsilon \bm{Z}) \left( \vartheta(\bm{Z}) \vartheta^{\top}(\bm{Z}) - \nabla^2 \vartheta(\bm{Z}) \right) \vartheta(\bm{Z}) d\bm{Z}  \\
    &= \mathbb{E} \left[ F_2(\bar{\bm{\theta}}_{t+1} + \epsilon \bm{Z}) \left( \vartheta(\bm{Z}) \vartheta^{\top}(\bm{Z}) - \nabla^2 \vartheta(\bm{Z}) \right) / \epsilon \right].
\end{aligned}
\end{equation}

Since these derivatives are continuous, we conclude that the loss function is differentiable, and the solution is well-defined.
% \end{proof}

%\end{proof}

%% file: main.bbl
\begin{thebibliography}{10}
\providecommand{\url}[1]{#1}
\csname url@samestyle\endcsname
\providecommand{\newblock}{\relax}
\providecommand{\bibinfo}[2]{#2}
\providecommand{\BIBentrySTDinterwordspacing}{\spaceskip=0pt\relax}
\providecommand{\BIBentryALTinterwordstretchfactor}{4}
\providecommand{\BIBentryALTinterwordspacing}{\spaceskip=\fontdimen2\font plus
\BIBentryALTinterwordstretchfactor\fontdimen3\font minus \fontdimen4\font\relax}
\providecommand{\BIBforeignlanguage}[2]{{%
\expandafter\ifx\csname l@#1\endcsname\relax
\typeout{** WARNING: IEEEtran.bst: No hyphenation pattern has been}%
\typeout{** loaded for the language `#1'. Using the pattern for}%
\typeout{** the default language instead.}%
\else
\language=\csname l@#1\endcsname
\fi
#2}}
\providecommand{\BIBdecl}{\relax}
\BIBdecl

\bibitem{CaliforniaStorage2024}
\BIBentryALTinterwordspacing
{Office of the Governor of California}, ``California exceeds another clean energy milestone,'' October 15 2024, accessed: 2025-02-28. [Online]. Available: \url{https://www.gov.ca.gov/2024/10/15/california-exceeds-another-clean-energy-milestone/}
\BIBentrySTDinterwordspacing

\bibitem{TexasStorage2025}
\BIBentryALTinterwordspacing
{Chron}, ``Texas is {U.S.} leader in renewable energy generation, says data,'' January 2025, accessed: 2025-02-28. [Online]. Available: \url{https://www.chron.com/news/houston-texas/article/texas-energy-solar-battery-california-20030398.php}
\BIBentrySTDinterwordspacing

\bibitem{zheng2023energy}
N.~Zheng, Q.~Xin, W.~Di, G.~Murtaugh, and B.~Xu, ``Energy storage state-of-charge market model,'' \emph{IEEE Transactions on Energy Markets, Policy and Regulation}, vol.~1, no.~1, pp. 11--22, 2023.

\bibitem{qin2023role}
X.~Qin, B.~Xu, I.~Lestas, Y.~Guo, and H.~Sun, ``The role of electricity market design for energy storage in cost-efficient decarbonization,'' \emph{Joule}, vol.~7, no.~6, pp. 1227--1240, 2023.

\bibitem{sioshansi2021energy}
R.~Sioshansi, P.~Denholm, J.~Arteaga, S.~Awara, S.~Bhattacharjee, A.~Botterud, W.~Cole, A.~Cortes, A.~De~Queiroz, J.~DeCarolis \emph{et~al.}, ``Energy-storage modeling: State-of-the-art and future research directions,'' \emph{IEEE transactions on power systems}, vol.~37, no.~2, pp. 860--875, 2021.

\bibitem{wang2017impact}
Y.~Wang, Y.~Dvorkin, R.~Fern{\'a}ndez-Blanco, B.~Xu, and D.~S. Kirschen, ``Impact of local transmission congestion on energy storage arbitrage opportunities,'' in \emph{2017 IEEE Power \& Energy Society General Meeting}.\hskip 1em plus 0.5em minus 0.4em\relax IEEE, 2017, pp. 1--5.

\bibitem{YAX24}
M.~Yi, S.~Alghumayjan, and B.~Xu, ``Perturbed decision-focused learning for modeling strategic energy storage,'' \emph{IEEE Transactions on Smart Grid}, vol.~16, no.~3, pp. 2574--2586, 2025.

\bibitem{SXL22}
L.~Sang, Y.~Xu, H.~Long, Q.~Hu, and H.~Sun, ``Electricity price prediction for energy storage system arbitrage: A decision-focused approach,'' \emph{IEEE Transactions on Smart Grid}, vol.~13, no.~4, pp. 2822--2832, 2022.

\bibitem{ruiz2012}
C.~Ruiz, A.~J. Conejo, and Y.~Smeers, ``Equilibria in an oligopolistic electricity pool with stepwise offer curves,'' \emph{IEEE Transactions on Power Systems}, vol.~27, no.~2, pp. 752--761, 2012.

\bibitem{hu2021}
B.~Hu, Y.~Gong, C.~Y. Chung, B.~F. Noble, and G.~Poelzer, ``Price-maker bidding and offering strategies for networked microgrids in day-ahead electricity markets,'' \emph{IEEE Transactions on Smart Grid}, vol.~12, no.~6, pp. 5201--5211, 2021.

\bibitem{chen2018}
S.~Chen, Q.~Chen, and Y.~Xu, ``Strategic bidding and compensation mechanism for a load aggregator with direct thermostat control capabilities,'' \emph{IEEE Transactions on Smart Grid}, vol.~9, no.~3, pp. 2327--2336, 2018.

\bibitem{zhang2020}
R.~Zhang, T.~Jiang, F.~Li, G.~Li, H.~Chen, and X.~Li, ``Coordinated bidding strategy of wind farms and power-to-gas facilities using a cooperative game approach,'' \emph{IEEE Transactions on Sustainable Energy}, vol.~11, no.~4, pp. 2545--2555, 2020.

\bibitem{shafiee2017}
S.~Shafiee, H.~Zareipour, A.~M. Knight, N.~Amjady, and B.~Mohammadi-Ivatloo, ``Risk-constrained bidding and offering strategy for a merchant compressed air energy storage plant,'' \emph{IEEE Transactions on Power Systems}, vol.~32, no.~2, pp. 946--957, 2017.

\bibitem{shafiee2019}
S.~Shafiee, H.~Zareipour, and A.~M. Knight, ``Developing bidding and offering curves of a price-maker energy storage facility based on robust optimization,'' \emph{IEEE Transactions on Smart Grid}, vol.~10, no.~1, pp. 650--660, 2019.

\bibitem{ZJX22}
N.~Zheng, J.~Jaworski, and B.~Xu, ``Arbitraging variable efficiency energy storage using analytical stochastic dynamic programming,'' \emph{IEEE Transactions on Power Systems}, vol.~37, no.~6, pp. 4785--4795, 2022.

\bibitem{BZX23}
Y.~Baker, N.~Zheng, and B.~Xu, ``Transferable energy storage bidder,'' \emph{IEEE Transactions on Power Systems}, 2023.

\bibitem{ADJ16}
K.~Abdulla, J.~De~Hoog, V.~Muenzel, F.~Suits, K.~Steer, A.~Wirth, and S.~Halgamuge, ``Optimal operation of energy storage systems considering forecasts and battery degradation,'' \emph{IEEE Transactions on Smart Grid}, vol.~9, no.~3, pp. 2086--2096, 2016.

\bibitem{CZZ17}
H.~Chitsaz, P.~Zamani-Dehkordi, H.~Zareipour, and P.~P. Parikh, ``Electricity price forecasting for operational scheduling of behind-the-meter storage systems,'' \emph{IEEE Transactions on Smart Grid}, vol.~9, no.~6, pp. 6612--6622, 2017.

\bibitem{EG22}
A.~N. Elmachtoub and P.~Grigas, ``Smart “predict, then optimize”,'' \emph{Management Science}, vol.~68, no.~1, pp. 9--26, 2022.

\bibitem{ELM20}
A.~N. Elmachtoub, J.~C.~N. Liang, and R.~McNellis, ``Decision trees for decision-making under the predict-then-optimize framework,'' in \emph{International conference on machine learning}.\hskip 1em plus 0.5em minus 0.4em\relax PMLR, 2020, pp. 2858--2867.

\bibitem{BJK24}
L.~Baty, K.~Jungel, P.~S. Klein, A.~Parmentier, and M.~Schiffer, ``Combinatorial optimization-enriched machine learning to solve the dynamic vehicle routing problem with time windows,'' \emph{Transportation Science}, 2024.

\bibitem{BBT20}
Q.~Berthet, M.~Blondel, O.~Teboul, M.~Cuturi, J.-P. Vert, and F.~Bach, ``Learning with differentiable pertubed optimizers,'' \emph{Advances in neural information processing systems}, vol.~33, pp. 9508--9519, 2020.

\bibitem{PPM20}
M.~V. Pogan{\v{c}}i{\'c}, A.~Paulus, V.~Musil, G.~Martius, and M.~Rolinek, ``Differentiation of blackbox combinatorial solvers,'' in \emph{International Conference on Learning Representations}, 2020.

\bibitem{DAK17}
P.~Donti, B.~Amos, and J.~Z. Kolter, ``Task-based end-to-end model learning in stochastic optimization,'' \emph{Advances in neural information processing systems}, vol.~30, 2017.

\bibitem{amos2017}
B.~Amos and J.~Z. Kolter, ``{OptNet}: Differentiable optimization as a layer in neural networks,'' in \emph{Proceedings of the 34th International Conference on Machine Learning}, 2017, pp. 136--145.

\bibitem{SX24}
Y.~Shi and B.~Xu, ``Demand-side price-responsive flexibility and baseline estimation through end-to-end learning,'' \emph{IET Renewable Power Generation}, vol.~18, no.~3, pp. 361--371, 2024.

\bibitem{BMN20}
M.~Blondel, A.~F. Martins, and V.~Niculae, ``Learning with fenchel-young losses,'' \emph{The Journal of Machine Learning Research}, vol.~21, no.~1, pp. 1314--1382, 2020.

\bibitem{PGS17}
A.~Paszke, S.~Gross, S.~Chintala, G.~Chanan, E.~Yang, Z.~DeVito, Z.~Lin, A.~Desmaison, L.~Antiga, and A.~Lerer, ``Automatic differentiation in {PyTorch},'' 2017.

\bibitem{PGM19}
A.~Paszke, S.~Gross, F.~Massa, A.~Lerer, J.~Bradbury, G.~Chanan, T.~Killeen, Z.~Lin, N.~Gimelshein, L.~Antiga \emph{et~al.}, ``Pytorch: An imperative style, high-performance deep learning library,'' \emph{Advances in neural information processing systems}, vol.~32, 2019.

\bibitem{TK22}
B.~Tang and E.~B. Khalil, ``{PyEPO}: A {PyTorch}-based end-to-end predict-then-optimize library for linear and integer programming,'' \emph{arXiv preprint arXiv:2206.14234}, 2022.

\bibitem{AAB19}
A.~Agrawal, B.~Amos, S.~Barratt, S.~Boyd, S.~Diamond, and J.~Z. Kolter, ``Differentiable convex optimization layers,'' \emph{Advances in neural information processing systems}, vol.~32, 2019.

\bibitem{SS16}
S.~Diamond and S.~Boyd, ``{CVXPY}: A {P}ython-embedded modeling language for convex optimization,'' \emph{Journal of Machine Learning Research}, vol.~17, no.~83, pp. 1--5, 2016.

\bibitem{C01}
M.~Carter, \emph{Foundations of mathematical economics}.\hskip 1em plus 0.5em minus 0.4em\relax MIT press, 2001.

\end{thebibliography}
